\documentclass[10pt,conference,compsocconf]{IEEEtran}

\usepackage{psfrag}
\usepackage{epsfig}
\usepackage{hyperref}
\usepackage{tabularx}
\usepackage{multirow}
\usepackage{subfigure}
\usepackage{multicol}
\usepackage{color}
\usepackage[active]{srcltx}
\usepackage{amsthm, amssymb}
\usepackage{amsmath,wasysym}
\usepackage{color}
\usepackage{cite}
\allowdisplaybreaks[2]          

\usepackage{amssymb} 

\usepackage{amsmath}
\allowdisplaybreaks[2]          

\usepackage[figure,boxed,noend]{algorithm2e}
\usepackage{graphicx}
\usepackage{subfigure}
\usepackage{hyperref}

\usepackage{float}
\usepackage[noend]{distribalgo}
\newfloat{algo}{htbp}{algo}
\floatname{algo}{Algorithm}

\usepackage[active]{srcltx}
\usepackage{amssymb} 

\newcommand{\protocol}{PoWerStore}
\newcommand{\mprotocol}{M-PoWerStore}
\newcommand{\complete}{\textsc{complete}}
\newcommand{\nonce}{N}
\newcommand{\hash}{\overline{N}}
\newcommand{\scheme}{storage technique}

\newtheorem{defn}{Definition}[section]
\newtheorem{la}[defn]{Lemma}

\newtheorem{theo}[defn]{Theorem}

\newenvironment{prooff}{\vspace{1ex}\noindent{\bf Proof:}\hspace{0.5em}}
	{\hfill\qed\vspace{1em}}

\title{Proofs of Writing for Efficient and Robust Storage}
\vspace{1em}
\author{Dan Dobre, Ghassan Karame, Wenting Li, Matthias Majuntke, Neeraj Suri and Marko Vukoli\'{c} \\
\{dan.dobre, ghassan.karame, wenting.li\}@neclab.eu,  \{majuntke, suri\}@cs.tu-darmstadt.de, vukolic@eurecom.fr}
\IEEEoverridecommandlockouts
\begin{document}

\maketitle
\thispagestyle{plain}
\pagestyle{plain}

\begin{abstract}



We present \protocol, the first efficient robust storage protocol that
achieves \emph{optimal latency} without using digital signatures.

\protocol's \emph{robustness} comprises  tolerating asynchrony, maximum
number of Byzantine storage servers, any number of Byzantine readers
and crash-faulty writers, and guaranteeing \emph{wait-freedom} and
\emph{linearizability} of read/write operations. Furthermore,
\protocol's \emph{efficiency} stems from combining \emph{lightweight
authentication}, \emph{erasure coding} and \emph{metadata write-backs}
where readers write-back only metadata to achieve linearizability.

At the heart of \protocol\ are \emph{Proofs of Writing (PoW)}: a novel
\scheme\ based on lightweight cryptography. PoW enable reads and
writes in the single-writer variant of \protocol\ to have latency of
2 rounds of communication between a client and storage servers in the
\emph{worst-case} (which we show optimal).

We further present and implement a multi-writer \protocol\ variant
featuring 3-round writes/reads where the third read round is invoked
only under active attacks, and show that it outperforms existing
robust storage protocols, including crash-tolerant ones.

\end{abstract}

\section{Introduction}

Byzantine fault-tolerant (BFT) distributed protocols have recently been attracting considerable research attention, due to their appealing promise of masking various system issues ranging from simple crashes, through software bugs and misconfigurations, all the way to intrusions and malware. However, there are many issues that render the use of existing BFT protocols questionable in practice; these include, e.g., weak guarantees under failures (e.g., \cite{SinghDMDR08,CWADM09,AmirCKL11}) or high cost in performance, deployment and maintenance when compared to \emph{crash-tolerant} protocols \cite{KR09}. This can help us derive the following requirements for the design of future BFT protocols:

\begin{itemize}
\item A BFT protocol should be \emph{robust}, i.e., it should tolerate actual Byzantine faults and  actual \emph{asynchrony} (modeling network outages) while maintaining correctness and providing sustainable progress even under \emph{worst-case} conditions that still meet the protocol assumptions. This requirement has often been neglected in BFT protocols that focus primarily on \emph{common}, failure-free operation modes (e.g., \cite{HendricksGR07,KotlaADCW09,GV10}).

\item A \emph{robust} protocol should be \emph{efficient} (e.g., \cite{GarciaRP11, VeroneseCBLV13}). We believe that the efficiency of a robust BFT protocol is best compared to the efficiency of its \emph{crash-tolerant} counterpart. Ideally, a robust protocol should not incur significant performance and resource cost penalty with respect to a crash-tolerant implementation, hence making the replacement of a crash-tolerant protocol a viable option.
\end{itemize}

We stand to the point that achieving these goals may require challenging existing approaches and revisiting the use of fundamental abstractions (such as cryptographic primitives).

In this paper, we focus on read/write \emph{storage} \cite{Lam86}, where a set of client (readers and writers) processes share data leveraging a set of storage
\emph{server} processes. Besides being fundamental, the read/write storage abstraction is also practically appealing given that it lies at the heart of the Key-Value Store (KVS) APIs --- a de-facto standard of modern cloud storage offerings.

In this context, we tackle the problem of developing a \emph{robust} and \emph{efficient} asynchronous distributed storage protocol. More specifically, storage \emph{robustness} implies \cite{ABND95}: \emph{(i)} \emph{wait-freedom} \cite{Her91}, i.e., read/write operations invoked by correct \emph{clients} always eventually return,
and \emph{(ii)} \emph{optimal resilience}, i.e., ensuring correctness despite the largest possible number $t$ of Byzantine server  failures; in the Byzantine model, this mandates using $3t+1$ servers \cite{MAD02}.

Our main contribution is \protocol, the first efficient robust storage protocol that
achieves \emph{optimal latency}, measured in \emph{maximum} (worst-case) number of communication \emph{rounds} between a client and storage servers, without using digital signatures. Perhaps surprisingly, the efficiency of \protocol\ does not come from sacrificing consistency; namely, \protocol\ ensures \emph{linearizability} \cite{HW90} (or \emph{atomicity} \cite{Lam86}) of read/write operations. In fact, the efficiency of \protocol{} stems from combining \emph{lightweight
authentication}, \emph{erasure coding} and \emph{metadata write-backs}
where readers write-back only metadata, avoiding much expensive data write-backs.

At the heart of \protocol\ is a novel data storage technique we call \emph{Proofs of Writing} (PoW). PoW are inspired by commitment schemes~\cite{Halevi1996};  PoW incorporate a 2-round write procedure, where the second round of write effectively serves to ``prove'' that the first round has actually been completed before it is exposed to a reader.
More specifically, PoW rely on the construction of a secure token that is \emph{committed} in the first write round, but can
only be verified after it is \emph{revealed} in the second round. Here, token security means that the adversary cannot predict nor forge the token
before the start of the second round. We construct PoW using cryptographic hash functions and efficient message authentication codes (MACs); in addition, we also propose an instantiation of PoW using
Shamir's secret sharing scheme \cite{Sha79}.

As a result, PoW allow \protocol\ to achieve 2-round read/write operations in the single writer (SW) case. This, in case of reads, matches the theoretical minimum for \emph{any} robust atomic storage implementation that supports arbitrary number of readers, \emph{including} crash-only implementations \cite{DGLV10}. Notably, \protocol\ is the first robust BFT storage protocol to achieve the 2-round latency of reading without using digital signatures.
On the other hand, we prove the 2-round write inherent, by showing a 2-round write lower bound for any robust BFT storage that features metadata write-backs using less than $4t+1$ storage servers.


In addition, \protocol\ employs \emph{erasure coding} at the client side to offload the storage servers and to reduce the amount of data transferred over the network. Besides being the first robust BFT storage protocol to feature metadata write-backs, \protocol\ is also the first robust BFT storage protocol to tolerate an unbounded number of Byzantine readers \cite{LR06} and unbounded number of writers' crash-faults.

Finally, while our SW variant of \protocol\ demonstrates the utility of PoW, for practical applications we propose a multi-writer (MW) variant of \protocol\ (referred to as \mprotocol). \mprotocol\ features 3-round writes and reads, where the third read round is invoked only under active attacks. \mprotocol{} also resists attacks specific to multi-writer setting that exhaust
the timestamp domain \cite{BD04}. We evaluate \mprotocol\ and demonstrate its superiority even with respect to existing crash-tolerant robust atomic storage implementations. Our results show that in typical settings, the peak throughput achieved by \mprotocol{} improves over existing crash-tolerant \cite{ABND95} and Byzantine-tolerant \cite{Phalanx} atomic storage implementations, by 50\% and 100\%, respectively.


The remainder of the paper is organized as follows. In Section~\ref{sec:model}, we outline our system model. In Section~\ref{sec:pow}, we introduce \protocol{} and we analyze its correctness.
In Section~\ref{sec:mpow}, we present the multi-writer variant of \protocol, \mprotocol{}. In Section~\ref{sec:implementation}, we evaluate an implementation of \mprotocol{}. In Section~\ref{sec:relwork}, we overview related work and we conclude the paper in Section~\ref{sec:conclusion}.

\section{System Model}
\label{sec:model}

We consider a distributed system that consists of three \emph{disjoint}
sets of processes: a set \emph{servers} of size $S=3t+1$, where $t$ is the failure threshold parameter, containing
processes $\{s_1, ..., s_S\}$; a collection \emph{writers} $w_1, w_2, ...$
and a collection \emph{readers} $r_1, r_2, ...$.
The collection \emph{clients} is the union of writers and readers.
We assume the \emph{data-centric} model \cite{CMD03,ChocklerGKV09,SMMK10} where every client may \emph{asynchronously} communicate with any server by message passing using point-to-point reliable
communication channels; however, servers cannot communicate
among each other, nor send messages to clients other
than in reply to clients' messages.

We further assume that each server $s_i$ pre-shares one symmetric group key with all writers
in the set $W$; in the following, we denote this key by $k_i$. In addition, we assume the existence of a cryptographic (i.e., one way and collision resistant) hash function $H(\cdot)$, and a secure message authentication function $MAC_k(\cdot)$, where $k$ is a $\lambda$-bit symmetric key.

We model processes as probabilistic I/O automata \cite{WSS94} where a distributed algorithm is a collection of such processes. Processes that follow the algorithm are called \emph{correct}. We assume that any number of readers exhibit \emph{Byzantine} \cite{LSP82} (or \emph{arbitrary} \cite{JCT98}) faults. Moreover, up to $t$ servers may be Byzantine. Byzantine processes can exhibit arbitrary behavior; however, we assume that Byzantine processes are computationally bounded and cannot break cryptographic hash functions or forge message authentication codes. Finally, any number of writers may fail by crashing.

We focus on a read/write storage abstraction \cite{Lam86} which exports two operations: \textsc{write}($v$), which stores value $v$ and \textsc{read}(), which returns the stored value. While every client may invoke the \textsc{read} operation, we assume that \textsc{write}s are invoked only by writers. We say that an
operation (invocation) $op$ is \emph{complete} if the client receives the \emph{response}, i.e., if the client returns from the invocation. We further assume that each correct client invokes at most one operation at a time (i.e., does not invoke the next operation
until it receives the response for the current operation).
We further assume that the initial value of a storage is a special value $\bot$, which is not a
valid input value for a write operation.

In any execution of an algorithm, we say that a complete
operation $op_1$ \emph{precedes} operation $op_2$ (or $op_2$ \emph{follows} $op_1$)
if the response of $op1$ precedes the invocation of
$op_2$ in that execution.

We focus on the strongest storage progress consistency and progress semantics, namely \emph{linearizability} \cite{HW90} (or \emph{atomicity} \cite{Lam86}) and \emph{wait-freedom} \cite{Her91}.
Wait-freedom states that if a correct client invokes an operation $op$, then
$op$ eventually completes. Linearizability gives an illusion that a complete operation $op$ by a correct client is executed instantly at some point in time between its invocation and response,  whereas the operations invoked by faulty clients appear either as complete or not invoked at all.

Finally, we measure the time-complexity of an atomic storage
implementation in terms of number of \emph{communication round-trips}
(or simply \emph{rounds}) between a client and servers. Intuitively, a \emph{round} consists of a client sending the message to (a subset of) servers and receiving replies. A formal definition can be found in, e.g.,~\cite{GNS09,DGLV10}.

\section{\protocol}\label{sec:pow}
In this section, we provide a detailed description of the \protocol\ protocol and we analyze its correctness. In addition, we show that \protocol{} exhibits
optimal (worst-case) latency in both \textsc{READ} and \textsc{WRITE} operations.

\subsection{Proofs of Writing}

At the heart of \protocol\ is a novel technique we call Proofs of Writing (PoW). Intuitively, PoW enable \protocol{} to complete in a 2-round \textsc{WRITE} procedure, where the second round of \textsc{WRITE} effectively serves to ``prove'' that the data is written in a quorum of servers (at least $S-t$) before it is exposed to a client. As such, PoW obviate the need for writing-back data, enabling efficient metadata write-backs and support for Byzantine readers.

PoW are inspired by commitment schemes~\cite{Halevi1996}; PoW consist of the construction of a secure token that can only be verified after the second round is invoked. Here, token security means that the adversary cannot predict nor forge the token before the start of the second round.
More specifically, our PoW implementation relies on the use of one-way collision-resistant functions seeded with pseudo-random input. We construct PoW as follows.

In the first round, the writer first generates a pseudo-random nonce and stores the hash of the nonce together with the data in a quorum of servers. The writer then discloses the nonce to the servers in the second round; this nonce provides sufficient \emph{proof} that the first round has actually completed. In fact, during the first round of a \textsc{READ} operation, the client collects the received nonces into a set and sends (writes-back) this set to the servers in the second round. The server then verifies the PoW by checking whether the received nonce matches the hash of the stored nonce. Note that since the writer keeps the nonce secret until the start of the second round, it is computationally infeasible for the adversary to acquire the nonce unless the first round of \textsc{WRITE} has been completed.

PoW are not restricted to the use of cryptographic hash functions. In Section~\ref{subsec:shamir}, we propose another possible instantiation of PoW using Shamir's secret sharing scheme \cite{Sha79}.

\subsection{Overview of \protocol}

In \protocol, the \textsc{WRITE} operation completes in two rounds, called \textsc{store} and \textsc{\complete}. Likewise, the \textsc{READ} performs in two rounds, called \textsc{collect} and \textsc{filter}. For the sake of convenience, each round $rnd \in \{$\textsc{store}, \textsc{\complete}, \textsc{collect}, \textsc{filter}$\}$ is wrapped by a procedure $rnd$. In each round $rnd$, the client sends a message of type $rnd$ to all servers. A round completes at the latest when the client receives messages of type $rnd\textsc{\_ack}$ from $S-t$ correct servers. The server maintains a variable $lc$ to store the timestamp of the last completed \textsc{WRITE}, and a variable $LC$, of the same structure, to maintain a set of timestamps written-back by clients. In addition, the server keeps a variable $Hist$ storing the history of the data written by the writer in the \textsc{store} round, indexed by timestamp.

\subsection{\textsc{WRITE} Implementation}

The \textsc{WRITE} implementation is given in Algorithm~\ref{alg:writer}. To write a value $V$, the writer\footnote{Recall that \protocol\ is a single-writer storage protocol.} increases its timestamp $ts$, computes a nonce $\nonce$ and its hash $\hash = H(\nonce)$, and invokes \textsc{store} with $ts$, $V$ and $\hash$. When the \textsc{store} procedure returns, the writer invokes \textsc{\complete} with $ts$ and $\nonce$. After \textsc{\complete} returns, the \textsc{WRITE} completes.

In \textsc{store}, the writer encodes $V$ into $S$ fragments $fr_i$ ($1 \leq i \leq S)$, such that $V$ can be recovered from any subset of $t+1$ fragments. Furthermore, the writer computes a cross-checksum $cc$ consisting of the hashes of each fragment. For each server $s_i$ $(1 \leq i \leq S)$, the writer sends a \textsc{store}$\langle ts, fr_i, cc, \hash \rangle$ message to $s_i$. On reception of such a message, the server writes $(fr_i, cc, \hash)$ into the history entry $Hist[ts]$ and replies to the writer.
After the writer receives $S-t$ replies from different servers, the \textsc{store} procedure returns, and the writer proceeds to \textsc{\complete}.

In \textsc{\complete}, the writer sends a \textsc{\complete}$\langle ts, \nonce \rangle$ message to all servers. Upon reception of such a message, the server changes the value of $lc$ to $(ts, \nonce)$ if $ts > lc.ts$ and replies to the writer. After the writer receives $S-t$ replies from different servers, the \textsc{\complete} procedure returns.

\begin{algo}[t]
\small
\newcounter{alg:client1:lines}

\begin{distribalgo}[1] \setcounter{ALC@line}{\value{alg:client1:lines}}
\smallskip
\INDENT {\textbf{Definitions:}}
\STATE $ts$ : structure $num$ initially $ts = ts_0 \triangleq 0$
\ENDINDENT

\setcounter{alg:client1:lines}{\value{ALC@line}}
\end{distribalgo}

\begin{tabular}{c}\hline\mbox{}\hspace{0.45\textwidth}\mbox{}\end{tabular}
\vspace{-2 em}
\begin{distribalgo}[1]  \setcounter{ALC@line}{\value{alg:client1:lines}}
\INDENT {\textbf{operation} \textsc{WRITE}$(V)$}
\STATE $ts \leftarrow ts + 1$ \label{alg:writer:timestamp}
\STATE $\nonce \leftarrow \{0,1\}^{\lambda}$
\STATE $\hash \leftarrow H(\nonce)$
\STATE $\textsc{store}(ts, \hash, V)$ \label{alg:writer:store}
\STATE $\textsc{\complete}(ts, \nonce)$ \label{alg:writer:complete}
\STATE return \textsc{ok}
\ENDINDENT

\medskip
\INDENT {\textbf{procedure} \textsc{store}($ts, V, \hash$)}
\STATE $\{fr_{1}, \dots, fr_{S}\} \leftarrow \mathrm{encode}(V,t+1,S)$
\STATE $cc \leftarrow [H(fr_{1}), \dots, H(fr_{S})]$
\STATE \textbf{for} $1 \leq i \leq S$ \textbf{do} send \textsc{store}$\langle ts, fr_i, cc, \hash\rangle$ to $s_i$
\STATE \textbf{wait for} \textsc{store\_ack}$\langle ts \rangle$ from $S-t$ servers
\ENDINDENT

\medskip
\INDENT {\textbf{procedure} \textsc{\complete}($ts, \nonce$)}
\STATE send \textsc{\complete}$\langle ts, \nonce \rangle$ to all servers
\STATE \textbf{wait for} \textsc{\complete\_ack}$\langle ts \rangle$ from $S-t$ servers
\ENDINDENT

\setcounter{alg:client1:lines}{\value{ALC@line}}

\end{distribalgo}

\begin{tabular}{c}\hline\mbox{}\hspace{0.45\textwidth}\mbox{}\end{tabular}

\caption{{Algorithm of the writer in \protocol.}}\label{alg:writer}
\end{algo}

\begin{algo*}[t]
\small
\centering
\begin{distribalgo}[1] \setcounter{ALC@line}{\value{alg:client1:lines}}
\smallskip
\INDENT {\textbf{Definitions:}}
\STATE $lc$ : structure $(ts, \nonce)$, initially $(ts_0, \textsc{null})$\hfill //last completed write
\STATE $LC$ : set of structure $(ts, \nonce)$, initially $\emptyset$\hfill     //set of written-back candidates
\STATE $Hist[\dots]$ : vector of $(fr,cc,\hash)$ indexed by $ts$, initially $Hist[ts_0] =  (\textsc{null},\textsc{null},\textsc{null})$
\ENDINDENT
\setcounter{alg:client1:lines}{\value{ALC@line}}
\end{distribalgo}
\begin{tabular}{c}\hline\mbox{}\hspace{0.97\textwidth}\mbox{}\end{tabular}
\begin{minipage}[t]{0.5\textwidth}
\begin{distribalgo}[1]  \setcounter{ALC@line}{\value{alg:client1:lines}}
\vspace{-1 em}
\INDENT {\textbf{upon} receiving \textsc{store}$\langle ts, fr, cc, \hash \rangle$ from the writer}
\STATE $Hist[ts] \leftarrow (fr, cc, \hash)$ \label{alg:server:hist-store}
\STATE send \textsc{store\_ack}$\langle ts \rangle$ to the writer
\ENDINDENT

\medskip
\INDENT {\textbf{upon} receiving \textsc{\complete}$\langle ts, \nonce \rangle$ from the writer}
\STATE  \textbf{if} {$ts > lc.ts$} \textbf{then} $lc \leftarrow (ts, \nonce)$ \label{alg:server:update-complete}
\STATE send \textsc{\complete\_ack}$\langle ts \rangle$ to the writer
\ENDINDENT

\medskip
\INDENT {\textbf{procedure} \textsc{gc}$()$}
\STATE $c_{hv} \leftarrow c_0$
\STATE $c_{hv} \leftarrow c \in LC : c $$=$$ \max(\{c \in LC  : \textsf{valid}(c)$\}) \label{alg:server:valid-check}
\STATE \textbf{if} {$c_{hv}.ts > lc.ts$} \textbf{then} $lc \leftarrow c_{hv}$ \label{alg:server:update-gc}
\STATE  $LC \leftarrow \{c \in LC : c.ts > lc.ts \wedge Hist[c.ts] = \textsc{null}\}$ \label{alg:server:gc}
\ENDINDENT

\setcounter{alg:client1:lines}{\value{ALC@line}}
\end{distribalgo}
\end{minipage}%
\hfill
\begin{minipage}[t]{0.5\textwidth}
\begin{distribalgo}[1]   \setcounter{ALC@line}{\value{alg:client1:lines}}
\vspace{-1 em}
\INDENT {\textbf{upon} receiving \textsc{collect}$\langle tsr \rangle$ from client $r$}
\STATE \textsc{gc}$()$
\STATE send \textsc{collect\_ack}$\langle tsr, LC \cup \{lc\} \rangle$ to client $r$
\ENDINDENT

\medskip

\INDENT {\textbf{upon} receiving \textsc{filter}$\langle tsr, C \rangle$ from client $r$}
\STATE $LC \leftarrow LC \cup C $ \hfill //write-back

\STATE $c_{hv} \leftarrow c_0$
\STATE $c_{hv} \leftarrow c \in C : c $$=$$ \max(\{c \in C: \textsf{valid}(c)\})$ \label{alg:server:valid-check2}
\STATE $(fr,cc) \leftarrow (Hist[c_{hv}.ts].fr, Hist[c_{hv}.ts].cc)$
\STATE send \textsc{filter\_ack}$\langle tsr, c_{hv}.ts, fr, cc \rangle$ to client $r$ \label{alg:server:ret}
\ENDINDENT

\medskip
\INDENT {\textbf{Predicates:}}
\STATE \textsf{valid}($c$) $\triangleq$ $(H(c.\nonce) = Hist[c.ts].\hash)$\label{alg:server:valid-pred}
\ENDINDENT

\setcounter{alg:client1:lines}{\value{ALC@line}}
\end{distribalgo}

\end{minipage}%

\medskip
\begin{tabular}{c}\hline\mbox{}\hspace{0.97\textwidth}\mbox{}\end{tabular}

\caption{{Algorithm of server $s_i$ in \protocol.}}\label{alg:server}
\end{algo*}

\subsection{\textsc{READ} Implementation}

The \textsc{READ} implementation is given in Algorithm~\ref{alg:reader}; it consists of the \textsc{collect} procedure followed by the \textsc{filter} procedure. In \textsc{collect}, the client reads the tuples $(ts, \nonce)$ included in the set $LC \cup \{lc\}$ at the server, and accumulates these tuples in a set $C$ together with the tuples read from other servers. We call such a tuple a \emph{candidate} and $C$ a \emph{candidate set}. Before responding to the client, the server garbage-collects old tuples in $LC$ using the $\textsc{gc}$ procedure. After the client receives candidates from $S-t$ different servers, \textsc{collect} returns.

In \textsc{filter}, the client submits $C$ to each server. Upon reception of $C$, the server performs a write-back of the candidates in $C$ (\emph{metadata write-back}). In addition, the server picks $c_{hv}$ as the candidate with the highest timestamp in $C$ such that \textsf{valid}($c_{hv}$) holds. The predicate \textsf{valid}($c$) holds if the server, based on the history, is able to verify the integrity of $c$ by checking that $H(c.\nonce)$ equals $Hist[c.ts].\hash$. The server then responds to the client with a message including the timestamp $c_{hv}.ts$, the fragment $Hist[c_{hv}.ts].fr$ and the cross-checksum $Hist[c_{hv}.ts].cc$. The client waits for the responses from servers until there is a candidate $c$ with the highest timestamp in $C$ such that \textsf{safe}($c$) holds, or until $C$ is empty, after which \textsc{collect} returns. The predicate \textsf{safe}($c$) holds if at least $t+1$ different servers $s_i$ have responded with timestamp $c.ts$, fragment $fr_i$ and cross-checksum $cc$ such that $H(fr_i) = cc[i]$. If $C \neq \emptyset$, the client selects the candidate with the highest timestamp $c \in C$ and restores value $V$ by decoding $V$ from the $t+1$ correct fragments received for $c$. Otherwise, the client sets $V$ to the initial value $\bot$. Finally, the \textsc{READ} returns $V$.

\begin{algo}[t]
\small

\begin{distribalgo}[1] \setcounter{ALC@line}{\value{alg:client1:lines}}
\smallskip
\INDENT {\textbf{Definitions:}}
\STATE $tsr$: $num$, initially $0$
\STATE $Q,R$: set of $pid$, initially $\emptyset$
\STATE $C$: set of $(ts, \nonce)$, initially $\emptyset$
\STATE $W[1 \dots S]$: vector of $(ts,fr,cc)$, initially $[]$
\ENDINDENT

\setcounter{alg:client1:lines}{\value{ALC@line}}
\end{distribalgo}
\begin{tabular}{c}\hline\mbox{}\hspace{0.45\textwidth}\mbox{}\end{tabular}
\vspace{-2 em}
\begin{distribalgo}[1]   \setcounter{ALC@line}{\value{alg:client1:lines}}

\INDENT {\textbf{operation} \textsc{READ}$()$}
\STATE $C,Q,R \leftarrow \emptyset$
\STATE $tsr \leftarrow tsr + 1$
\STATE $C \leftarrow \textsc{collect}(tsr)$
\STATE $C \leftarrow \textsc{filter}(tsr, C)$
\IF {$C \neq \emptyset$}
\STATE $c \leftarrow c' \in C: \textsf{highcand}(c') \wedge \textsf{safe}(c')$~\label{alg:reader:select}
\STATE $V \leftarrow \textsc{restore}(c.ts)$ ~\label{alg:reader:restore}
\ENDIF
\STATE \textbf{else} $V \leftarrow \bot$
\STATE return $V$
\ENDINDENT

\medskip
\INDENT {\textbf{procedure} \textsc{collect}($tsr$)}
\STATE send \textsc{collect}$\langle tsr \rangle$ to all servers
\STATE \textbf{wait until} $|Q| \geq S-t$
\STATE return $C$
\ENDINDENT

\smallskip
\INDENT {\textbf{upon} receiving \textsc{collect\_ack}$\langle tsr, C_i\rangle$ from server $s_i$}
\STATE $Q \leftarrow Q \cup \{ i \}$
\STATE $C \leftarrow C \cup \{c \in C_i : c.ts > ts_0\}$
\ENDINDENT

\medskip
\INDENT {\textbf{procedure} \textsc{filter}($tsr, C$)}
\STATE send \textsc{filter}$\langle tsr, C \rangle$ to all servers
\STATE \textbf{wait until} $|R| \geq S-t\ \wedge$\\ $((\exists c \in C: \textsf{highcand}(c) \wedge \textsf{safe}(c)) \vee C = \emptyset)$
\STATE return $C$
\ENDINDENT

\smallskip
\INDENT {\textbf{upon} receiving \textsc{filter\_ack}$\langle tsr, ts, fr, cc\rangle$ from server $s_i$}
\STATE $R \leftarrow R \cup \{ i \}$; $W[i] \leftarrow (ts,fr,cc)$
\STATE $C \leftarrow C \setminus \{c \in C: \textsf{invalid}(c) \}$
\ENDINDENT

\medskip
\INDENT {\textbf{procedure} \textsc{restore}($ts$)}
\STATE $cc \leftarrow cc'$ s.t. $\exists R' \subseteq R : |R'| \geq t+1\ \bigwedge$\\ $(\forall i \in R' : W[i].ts = ts \wedge W[i].cc = cc')$ \label{alg:reader:cc}
\STATE $F \leftarrow \{W[i].fr : i $$\in$$ R \wedge W[i].ts $$=$$ ts \wedge H(W[i].fr) $$=$$ cc[i]\}$ \label{alg:reader:fr}
\STATE $V \leftarrow \mathrm{decode}(F, t+1, S)$ \label{alg:reader:decode}
\STATE return $V$
\ENDINDENT

\medskip
\INDENT {\textbf{Predicates:}}
\STATE \textsf{highcand}($c$) $\triangleq$ ($c.ts = \max(\{c'.ts : c' \in C\}))$
\smallskip
\STATE \textsf{safe}($c$) $\triangleq$ $\exists R' \subseteq R: |R'| \geq t+1 \ \bigwedge$\\ \label{alg:reader:safe}
$(\forall i \in R': W[i].ts = c.ts)\ \bigwedge$\\
$(\forall i,j \in R'$$:$$W[i].cc $$=$$ W[j].cc \wedge H(W[i].fr) $$=$$ W[j].cc[i])$
\smallskip
\STATE \textsf{invalid}($c$) $\triangleq$ $|\{i \in R: W[i].ts < c.ts \}| \geq S-t$
\ENDINDENT

\setcounter{alg:client1:lines}{\value{ALC@line}}
\end{distribalgo}
\begin{tabular}{c}\hline\mbox{}\hspace{0.45\textwidth}\mbox{}\end{tabular}
\caption{{Algorithm of client $r$ in \protocol.}}\label{alg:reader}
\end{algo}

\subsection{Correctness Sketch} \label{sec:pow-sketch}

In what follows, we show that \protocol{} satisfies linearizability and wait-freedom in all \textsc{filter}. Due to lack of space, we refer the readers to the Appendix for a detailed proof of \protocol's correctness.


We first explain why linearizability is satisfied by arguing that if a \textsc{READ} follows after a completed \textsc{WRITE}($V$) (resp. a completed \textsc{READ} that returns $V$) then the \textsc{READ} does not return a value older than $V$.

\paragraph{\textsc{READ}/\textsc{WRITE} Linearizability}
Suppose a \textsc{READ} $rd$ by a correct client follows after a completed \textsc{WRITE}($V$). If $ts$ is the timestamp of \textsc{WRITE}$(V)$, we argue that $rd$ does not select a candidate with a timestamp lower than $ts$. Since a correct server never changes $lc$ to a candidate with a lower timestamp, after \textsc{WRITE}$(V)$ completed, $t+1$ correct servers hold a valid candidate with timestamp $ts$ or greater in $lc$. Hence, during \textsc{collect} in $rd$, a valid candidate $c$ with timestamp $ts$ or greater is included in $C$. Since $c$ is valid, at least $t+1$ correct servers hold history entries matching $c$ and none of them respond with a timestamp lower than $c.ts$. Consequently, at most $2t$ timestamps received by the client in \textsc{filter} are lower than $c.ts$ and thus $c$ is never excluded from $C$. By Algorithm~\ref{alg:reader}, $rd$ does not select a candidate with timestamp lower than $c.ts \geq ts$.

\paragraph{\textsc{READ} Linearizability}
Suppose a \textsc{READ} $rd'$ by a correct client follows after $rd$. If $c$ is the candidate selected in $rd$, we argue that $rd'$ does not select a candidate with a timestamp lower than $c.ts$.
By the time $rd$ completes, $t+1$ correct servers hold $c$ in $LC$. According to Algorithm~\ref{alg:server}, if a correct server excludes $c$ from $LC$, then the server changed $lc$ to a valid candidate with timestamp $c.ts$ or greater. As such, $t+1$ correct servers hold in $LC \cup \{ lc \}$ a valid candidate with timestamp $c.ts$ or greater and during \textsc{collect} in $rd'$, a valid candidate $c'$ with timestamp $c.ts$ or greater is included in $C$. By applying the same arguments as above, $rd$ does not select a candidate with timestamp lower than $c'.ts \geq c.ts$.

We now show that \protocol{} satisfies wait-freedom; here, we argue that the \textsc{READ} does not block in \textsc{filter} while waiting for the candidate $c$ with the highest timestamp in $C$ to become \textsf{safe}($c$) or $C$ to become empty.

\paragraph{Wait-freedom}
Suppose by contradiction that $rd$ blocks during \textsc{filter} after receiving \textsc{filter\_ack} messages from all correct servers. We distinguish two cases: (Case 1) $C$ contains a valid candidate and (Case 2) $C$ contains no valid candidate.

\begin{itemize}
\item \textbf{Case 1:} Let $c$ be the valid candidate with the highest timestamp in $C$. As $c$ is valid, at least $t+1$ correct servers hold history entries matching $c$. Since no valid candidate in $C$ has a higher timestamp than $c.ts$, these $t+1$ correct servers responded with timestamp $c.ts$, corresponding erasure coded fragment $fr_i$ and cross-checksum $cc$ such that $H(fr_i) = cc[i]$. Therefore, $c$ is \textsf{safe}$(c)$. Furthermore, all correct servers (at least $S-t$) responded with timestamps at most $c.ts$. Hence, every candidate $c' \in C$ such that $c'.ts > c.ts$ becomes \textsf{invalid}$(c')$  and is excluded from $C$. As such, $c$ is also \textsf{highcand}$(c)$ and we conclude that $rd$ does not block.
\item \textbf{Case 2:} Since none of the candidates in $C$ is valid, all correct servers (at least $S-t$) responded with timestamp $ts_0$, which is lower than any candidate timestamp. As such,  every candidate $c \in C$  becomes \textsf{invalid}$(c)$ is excluded from $C$. We therefore conclude that $rd$ does not block.
\end{itemize}

\subsection{PoW based on Shamir's Secret Sharing Scheme}\label{subsec:shamir}

In what follows, we propose an alternative construction of PoW based on Shamir's secret sharing scheme~\cite{Sha79}.
Here, the writer constructs a polynomial $P(\cdot)$ of degree $t$ with coefficients $\{\alpha_{t}, \dots, \alpha_{0}\}$ chosen at random from ${\cal Z}_{q}$, where $q$ is a public parameter. That is,
$P(x) = \sum_{j=0}^{j=t} \alpha_j x^j$.

The writer then constructs the PoW as follows: for each server $s_i$, the writer picks a random point $x_i$, and constructs the share $(x_{i}, P_i)$, where $P_i = P(x_i)$. As such, the writer constructs $S$ different shares, one for each server, and sends a \textsc{store}$\langle ts, fr_i, cc, x_{i}, P_i \rangle$ message to each server $s_i$ over a \emph{confidential channel}.
Upon reception of such a message, server $s_i$ stores $(fr_i, cc, x_i, P_i)$ in $Hist[ts]$. Note that since there are at most $t$ Byzantine servers, these servers cannot reconstruct the polynomial $P(\cdot)$ from their shares, even if they collude. In the \textsc{\complete} round, the writer reveals the polynomial $P(\cdot)$ in a \textsc{\complete}$\langle ts, P(\cdot) \rangle$ message. This enables a client to determine for a candidate $c$ = $(ts, P(\cdot))$ that the corresponding \textsc{store} round completed by checking that $t+1$ servers $s_i$ stored $(x_{i}, P(x_{i}))$, without the servers revealing their share. For this purpose, the \textsf{valid} predicate at server $s_i$ changes to $\textsf{valid}(c) \triangleq (c.P(Hist[c.ts].x_i) = Hist[c.ts].P_i)$.


By relying on randomly chosen $x_i$, and the fact that correct servers never divulge their share, our construction prevents an adversary from fabricating a \emph{partially corrupted} polynomial after the disclosure of $P(\cdot)$. To see why, note that with the knowledge of $P(\cdot)$ and $x_{i}$ held by a correct server $s_i$, the adversary could fabricate a partially corrupted polynomial $\hat{P(\cdot)}$ that intersects with $P(\cdot)$ only at $x_{i}$ (i.e., $P(x_i) = \hat{P(x_i)}$). This would lead to a situation in which a candidate $c$ is neither \textsf{safe}$(c)$ nor \textsf{invalid}$(c)$ and thus, the \textsc{READ} would block.

We point that, unlike the solution based on hash functions, this construction provides information-theoretic guarantees for the PoW~\cite{LCAA07,AAB07} during the \textsc{store} round.

\subsection{Optimality of \protocol{}}
In this section, we prove that \protocol\ features optimal \textsc{WRITE} latency, by showing that writing in two rounds is necessary. We start by giving some informal definitions.

A distributed algorithm $A$ is a collection of automata \cite{LT89}, where automaton $A_p$ is assigned to process $p$. Computation proceeds in steps of $A$ and a \emph{run} is an infinite sequence of steps of $A$. A \emph{partial run} is a finite prefix of some run. We say that a (partial) run $r$ \emph{extends} some partial run $pr$ if $pr$ is a \emph{prefix} of $r$.

We say that an implementation of linearizable storage is \emph{selfish}, if readers write-back only metadata instead of the full value. Intuitively the readers are selfish because they do not help the writer complete a write. For a formal definition, we refer the readers to~\cite{FL03}. Furthermore, we say that a \textsc{WRITE} operation is \emph{fast} if it completes in a single round. We now proceed to proving the main result. 


\begin{theo} \label{theo:lb}
There is no \emph{fast} \textsc{WRITE} implementation $I$ of a multi-reader \emph{selfish} robust linearizable storage that makes use of less than $4t+1$ servers.
\end{theo}
\noindent \textbf{Preliminaries}.
We prove Theorem~\ref{theo:lb} by contradiction assuming at most $4t$ servers. We partition the set of servers into four distinct subsets (we call \emph{blocks}), denoted by $T_1$, $T_2$, $T_3$ each of size exactly $t$, and $T_4$ of size at least $1$ and at most $t$. Without loss of generality we assume that each block contains at least one server. We say that an operation $op$ \emph{skips} a block $T_i$, ($1 \leq i \leq 4$) when all messages by $op$ to $T_i$ are delayed indefinitely (due to asynchrony) and all other blocks $T_j$ receive all messages by $op$ and reply.

\begin{figure}[tbp]
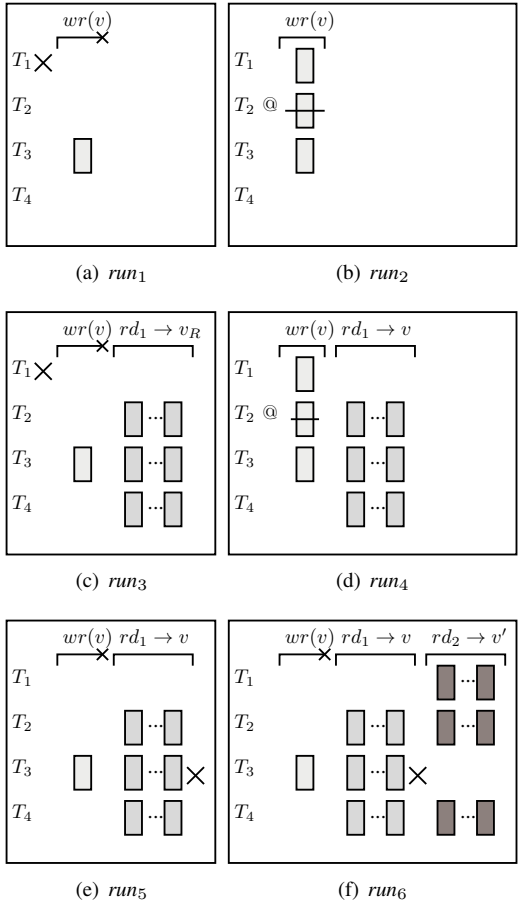

 \begin{center}
 \subfigure[$\emph{run}_1$]{\label{Fig:r1}\includegraphics[scale=0.75]{atomic_r1.mps}}
 \subfigure[$\emph{run}_2$]{\label{Fig:r2}\includegraphics[scale=0.75]{atomic_r2.mps}}

 \subfigure[$\emph{run}_3$]{\label{Fig:r3}\includegraphics[scale=0.75]{atomic_r3.mps}}
 \subfigure[$\emph{run}_4$]{\label{Fig:r4}\includegraphics[scale=0.75]{atomic_r4.mps}}

  \subfigure[$\emph{run}_5$]{\label{Fig:r5}\includegraphics[scale=0.75]{atomic_r5.mps}}
  \subfigure[$\emph{run}_6$]{\label{Fig:r6}\includegraphics[scale=0.75]{atomic_r6.mps}}

 \end{center}
 \vspace{-0.5cm}
 \caption{Sketch of the runs used in the proof of Theorem \ref{theo:lb}.}
\label{fig:proof_lb}
\end{figure}

\begin{prooff}
We construct a series of runs of a linearizable implementation $I$ towards a partial run that violates linearizability.
\begin{itemize}
\item Let $run_1$ be the partial run in which all servers are correct except $T_1$ which crashed at the beginning of $run_1$. Let $wr$ be the operation invoked by the writer $w$ to write a value $v \neq \bot$ in the storage. The \textsc{WRITE} $wr$ is the only operation invoked in $run_1$ and $w$ crashes after writing $v$ to $T_3$. Hence, $wr$ skips blocks $T_1$, $T_2$ and $T_4$.
\item Let $run'_1$ be the partial run in which all servers are correct except $T_4$, which crashed at the beginning of $run'_1$. In $run'_1$, $w$ is correct and $wr$ completes by writing $v$ to all blocks except $T_4$, which it skips.
\item Let $run_2$ be the partial run similar to $run'_1$, in which all servers except $T_2$ are correct, but due to asynchrony, all messages from $w$ to $T_4$ are delayed. Like in $run'_1$, $wr$ completes by writing $v$ to all servers except $T_4$, which it skips. To see why, note that $wr$ cannot distinguish $run_2$ from $run'_1$. After $wr$ completes, $T_2$ fails Byzantine by reverting its memory to the initial state.
\item Let $run_3$ extend $run_1$ by appending a complete \textsc{READ} $rd_1$ invoked by $r_1$. By our assumption, $I$ is wait-free. As such, $rd_1$ completes by skipping $T_1$ (because $T_1$ crashed) and returns (after a finite number of rounds) a value $v_R$.
\item Let $run_4$ extend $run_2$ by appending $rd_1$. In $run_4$, all servers except $T_2$ are correct, but due to asynchrony all messages from $r_1$ to $T_1$ are delayed indefinitely. Moreover, since $T_2$ reverted its memory to the initial state, $v$ is held only by $T_3$. Note that $r_1$ cannot distinguish $run_4$ from $run_3$ in which $T_1$ has crashed. As such, $rd_1$ completes by skipping $T_1$ and returns $v_R$. By linearizability, $v_R = v$.
\item Let $run_5$ be similar to $run_3$ in which all servers except $T_3$ are correct but, due to asynchrony, all messages from $r_1$ to $T_1$ are delayed. Note that $r_1$ cannot distinguish $run_5$ from $run_3$. As such, $rd_1$ returns $v_R$ in $run_5$, and by $run_4$, $v_R = v$. After $rd_1$ completes, $T_3$ fails by crashing.
\item Let $run_6$ extend $run_5$ by appending a \textsc{READ} $rd_2$ invoked by $r_2$ that completes by returning $v'$. Note that in $run_5$, \emph{(i)} $T_3$ is the only server to which $v$ was written, \emph{(ii)} $rd_1$ did not write-back $v$ (to any other server) before returning $v$, and \emph{(iii)} $T_3$ crashed before $rd_2$ is invoked. As such, $rd_2$ does not find $v$ in any server and hence $v' \neq v$, violating linearizability.
\end{itemize}
\vspace{-1.5 em}
\end{prooff}

Notice that Theorem~\ref{theo:lb} applies even to implementations relying on self-verifying data and/or not tolerating Byzantine readers. Furthermore, the proof extends to crash-tolerant storage when deleting the Byzantine block $T_2$ in the above proof; the result being that there is no selfish implementation of a multi-reader crash-tolerant linearizable storage with less than $3t+1$ servers in which every \textsc{WRITE} is fast.

\section{\mprotocol}\label{sec:mpow}

The \protocol\ protocol, as presented in Section~\ref{sec:pow}, has a drawback in having potentially very large candidate sets that servers send to clients and that clients write-back to servers. As a result, a malicious adversary can exploit the fact that in \protocol\ candidate sets can (theoretically) grow without bounds and mount
a denial of service (DoS) attack by fabricating huge sets of bogus candidates. While this attack can be contained by a robust implementation of the point-to-point channel assumption using, e.g., a separate pair of network cards for each channel (in the vein of \cite{CWADM09}), this may impact practicality of \protocol. To rectify this issue, and for practical applications, we propose a multi-writer variant of our protocol called \mprotocol.

\subsection{Overview}

\mprotocol\ (Algorithms~\ref{alg2:writer},~\ref{alg2:server} and~\ref{alg2:reader}) supports an unbounded number of clients. In addition, \mprotocol\ features optimal \textsc{READ} latency of two rounds in the \emph{common case}, where no process is Byzantine. Outside the common case, under active attacks, \mprotocol\ gracefully degrades to guarantee reading in at most three rounds. The \textsc{WRITE} has a latency of three rounds, featuring non-skipping timestamps \cite{BD04}, which prevents attacks specific to multi-writer setting that exhaust the timestamp domain.

The main difference between \mprotocol\ and \protocol\ is that, here, servers keep a single written-back candidate instead of a set. To this end, it is crucial that servers are able to determine the validity of a written-back candidate without consulting the history. For this purpose, we enhance our original PoW scheme by extending the candidate with message authentication codes (MACs) to authenticate the timestamp and the nonce's hash, one for each server, using the corresponding group key. As such, a valid MAC proves to a server that the timestamp has been issued by a writer in \complete, and thus, constitutes a PoW that a server can obtain even without the corresponding history entry. Note that in case of a candidate incorporating corrupted MACs, servers might disagree about the validity of a written-back candidate. Hence, a correct client might not be able to write-back a candidate to $t+1$ correct servers as needed. To solve this issue, \mprotocol\ "pre-writes" the MACs in the \textsc{store} round, enabling the client to repair the broken MACs of a selected candidate. We point out that the adversary cannot forge the MACs (and therefore bypass the PoW) before the start of the \complete, since the constructed MACs, besides the timestamp, also include the nonce's hash.

To support multiple-writers, \textsc{WRITE} in \mprotocol\ comprises an additional clock synchronization round, called \textsc{clock}, which is prepended to \textsc{store}. The \textsc{READ} performs an additional round called \textsc{repair}, which is appended to \textsc{collect}. The purpose of \textsc{repair} is to repair and write-back candidates if necessary, and is invoked only under active attack by a malicious adversary that actually corrupts candidates.

Similarly to \protocol, the server maintains the variable $Hist$ to store the history of the data written by the writer in the \textsc{store} round, indexed by timestamp. In addition, the server keeps the variable $lc$ to store the timestamp of the last completed write.

\subsection{\textsc{WRITE} Implementation}
The full \textsc{WRITE} implementation is given in Algorithm~\ref{alg2:writer}. In the following, we simply highlight the differences to \protocol.

As outlined before, \mprotocol\ is resilient to the adversary skipping timestamps. This is achieved by having the writer authenticate the timestamp of a \textsc{WRITE} with a key $k_W$ shared among the writers. Note that such a shared key can be obtained by combining the different group keys; for instance, $k_W \leftarrow H(k_1 || k_2 || \dots)$.

To obtain a timestamp, in the \textsc{clock} procedure, the writer retrieves the timestamp (held in variable $lc$) from a quorum of $S-t$ servers and picks the highest timestamp $ts$ with a valid MAC. Then, the writer increases $ts$ and computes a MAC for $ts$ using $k_W$. Finally, \textsc{clock} returns $ts$.

To write a value $V$, the writer, \emph{(i)} obtains a timestamp $ts$ from the \textsc{clock} procedure, \emph{(ii)} authenticates $ts$ and the nonce's hash $\hash$ by a vector of MACs $vec$, with one entry for each server $s_i$ using group key $k_i$, and \emph{(iii)} stores $vec$ both in \textsc{store} and \complete. Upon reception of a \textsc{store}$\langle fr_i, cc, \hash, vec \rangle$ message, the server writes the tuple $(fr_i, cc, \hash, vec)$ into the history entry $Hist[ts]$. Upon reception of a \complete$\langle ts, \nonce, vec \rangle$ message, the server changes the value of $lc$ to $(ts, \nonce, vec)$ if $ts > lc.ts$.

\begin{algo}[t]
\small

\begin{distribalgo}[1] \setcounter{ALC@line}{\value{alg:client1:lines}}
\smallskip
\INDENT {\textbf{Definitions:}}
\STATE $Q$: set of $pid$, (process id) initially $\emptyset$
\STATE $ts$: structure $(num, pid, \textit{MAC}_{\{k_W\}}(num||pid))$,\\ initially $ts_0$ $\triangleq$ $(0,0,\textsc{null})$
\ENDINDENT
\setcounter{alg:client1:lines}{\value{ALC@line}}
\end{distribalgo}
\begin{tabular}{c}\hline\mbox{}\hspace{0.45\textwidth}\mbox{}\end{tabular}
\vspace{-2 em}
\begin{distribalgo}[1]  \setcounter{ALC@line}{\value{alg:client1:lines}}
\INDENT {\textbf{operation} \textsc{WRITE}$(V)$}
\STATE $Q \leftarrow \emptyset$
\STATE $ts \leftarrow \textsc{clock}()$ \label{alg2:writer:clock}
\STATE $\nonce \leftarrow \{0,1\}^{\lambda}$
\STATE $\hash \leftarrow H(\nonce)$
\STATE $vec \leftarrow [\textit{MAC}_{\{k_{i}\}}(ts||\hash)_{1 \leq i\leq S}]$
\STATE $\textsc{store}(ts, V, \hash, vec)$ \label{alg2:writer:store}
\STATE $\complete(ts, \nonce, vec)$ \label{alg2:writer:complete}
\STATE return \textsc{ok}
\ENDINDENT

\medskip
\INDENT {\textbf{procedure} \textsc{clock}()}
\STATE send \textsc{clock}$\langle ts \rangle$ to all servers
\STATE \textbf{wait until} $|Q| \geq S-t$
\STATE $ts.num \leftarrow ts.num + 1$
\STATE $ts \leftarrow(ts.num, w, \textit{MAC}_{\{k_W\}}(ts.num||w))$
\STATE return $ts$
\ENDINDENT

\smallskip
\INDENT {\textbf{upon} receiving \textsc{clock\_ack}$\langle ts, ts_i \rangle$ from server $s_i$}
\STATE $Q \leftarrow Q \cup \{i\}$
\STATE \textbf{if} $ts_i > ts \wedge \mathrm{verify}(ts_i, k_W)$ \textbf{then} $ts \leftarrow ts_i$ \label{alg2:writer:ts-integrity}
\ENDINDENT

\medskip
\INDENT {\textbf{procedure} \textsc{store}($ts, V, \hash, vec$)}
\STATE $\{fr_{1}, \dots, fr_{S}\} \leftarrow \mathrm{encode}(V,t+1,S)$
\STATE $cc \leftarrow [H(fr_{1}), \dots, H(fr_{S})]$
\STATE \textbf{foreach} server $s_i$ send \textsc{store}$\langle ts, fr_i, cc, \hash, vec\rangle$ to $s_i$
\STATE \textbf{wait for} \textsc{store\_ack}$\langle ts \rangle$ from $S-t$ servers
\ENDINDENT

\medskip
\INDENT {\textbf{procedure} \complete($ts, \nonce, vec$)}
\STATE send \complete$\langle ts, \nonce, vec \rangle$ to all servers
\STATE \textbf{wait for} \textsc{\complete\_ack}$\langle ts \rangle$ from $S-t$ servers
\ENDINDENT

\setcounter{alg:client1:lines}{\value{ALC@line}}

\end{distribalgo}
\begin{tabular}{c}\hline\mbox{}\hspace{0.45\textwidth}\mbox{}\end{tabular}
\caption{{Algorithm of writer $w$ in \mprotocol.}}\label{alg2:writer}
\end{algo}

\begin{algo*}[t]
\small
\centering
\begin{distribalgo}[1] \setcounter{ALC@line}{\value{alg:client1:lines}}
\smallskip
\INDENT {\textbf{Definitions:}}
\STATE $lc$: structure $(ts, \nonce, vec)$, initially $c_0 \triangleq (ts_0, \textsc{null}, \textsc{null})$ \hfill
\STATE $Hist[\dots]$: vector of $(fr, cc, \hash, vec)$ indexed by $ts$, initially $Hist[ts_0] =  (\textsc{null}, \textsc{null}, \textsc{null}, \textsc{null})$
\ENDINDENT
\end{distribalgo}


\begin{tabular}{c}\hline\mbox{}\hspace{0.97\textwidth}\mbox{}\end{tabular}
\begin{minipage}[t]{0.44\textwidth}
\begin{distribalgo}[1]  \setcounter{ALC@line}{\value{alg:client1:lines}}
\vspace{-1 em}
\INDENT {\textbf{upon} receiving \textsc{clock}$\langle ts \rangle$ from writer $w$}
\STATE send \textsc{clock\_ack}$\langle ts, lc.ts \rangle$ to writer $w$
\ENDINDENT

\medskip
\INDENT {\textbf{upon} receiving \textsc{store}$\langle ts, fr, cc, \hash, vec \rangle$\\ from writer $w$}
\STATE $Hist[ts] \leftarrow (fr, cc, \hash, vec)$
\STATE send \textsc{store\_ack}$\langle ts \rangle$ to writer $w$
\ENDINDENT

\medskip
\INDENT {\textbf{upon} receiving \complete$\langle ts, \nonce, vec \rangle$ from writer $w$}
\STATE  \textbf{if} {$ts > lc.ts$} \textbf{then} $lc \leftarrow (ts, \nonce, vec)$\label{alg2:server:update-complete}
\STATE send \textsc{\complete\_ack}$\langle ts\rangle$ to writer $w$
\ENDINDENT

\medskip
\INDENT {\textbf{upon} receiving \textsc{collect}$\langle tsr \rangle$ from client $r$}
\STATE send \textsc{collect\_ack}$\langle tsr, lc \rangle$ to client $r$
\ENDINDENT

\setcounter{alg:client1:lines}{\value{ALC@line}}
\end{distribalgo}
\end{minipage}%
\hfill
\begin{minipage}[t]{0.56\textwidth}
\begin{distribalgo}[1]   \setcounter{ALC@line}{\value{alg:client1:lines}}
\vspace{-1 em}
\INDENT {\textbf{upon} receiving \textsc{filter}$\langle tsr, C \rangle$ from client $r$}
\STATE $c_{wb}, c_{rt} \leftarrow c_0$
\STATE $c_{wb} \leftarrow c \in C: c$$=$$\max(\{c \in C: \textsf{valid}(c)\})$ \label{alg2:server:valid-filter}
\STATE \textbf{if} $c_{wb}.ts > lc.ts$ \textbf{then} $lc \leftarrow c_{wb} $\label{alg2:server:update-filter} \hfill //write-back
\smallskip
\STATE $c_{rt} \leftarrow c \in C: c$$=$$\max(\{c \in C: \textsf{validByHist}(c)\})$ \label{alg2:server:valid-nonce}
\STATE $(fr, cc, vec) \leftarrow (Hist[c_{rt}.ts].fr, Hist[c_{rt}.ts].cc, Hist[c_{rt}.ts].vec)$
\STATE send \textsc{filter\_ack}$\langle tsr, c_{rt}.ts, fr, cc, vec \rangle$ to client $r$ \label{alg2:server:ret}
\ENDINDENT

\medskip
\INDENT {\textbf{upon} receiving \textsc{repair}$\langle tsr, c \rangle$ from client $r$}
\STATE \textbf{if} $c.ts > lc.ts \wedge \textsf{valid}(c)$ \textbf{then} $lc \leftarrow c$ \label{alg2:server:update-repair}\label{alg2:server:valid-repair}
\STATE send \textsc{repair\_ack}$\langle tsr\rangle$ to client $r$
\ENDINDENT

\medskip
\INDENT {\textbf{Predicates:}}
\STATE \textsf{valid}($c$) $\triangleq$ ($\textsf{validByHist}(c)$ $\vee$ $\mathrm{verify}(c.vec[i], c.ts, H(c.\nonce), k_i)$) \label{alg2:server:valid-pred}
\STATE \textsf{validByHist}($c$) $\triangleq$ ($H(c.\nonce) = Hist[c.ts].\hash$)
\ENDINDENT

\setcounter{alg:client1:lines}{\value{ALC@line}}
\end{distribalgo}

\end{minipage}%

\begin{tabular}{c}\hline\mbox{}\hspace{0.97\textwidth}\mbox{}\end{tabular}
\vspace{-1 em}
\caption{{Algorithm of server $s_i$ in \mprotocol.}}\label{alg2:server}
\end{algo*}

\subsection{\textsc{READ} Implementation}
The full \textsc{READ} implementation is given in Algorithm~\ref{alg:reader}.
The \textsc{READ} consists of three consecutive rounds, \textsc{collect}, \textsc{filter} and \textsc{repair}. In \textsc{collect}, a client reads the candidate triple $(ts, K, vec)$ stored in variable $lc$ in the server, and inserts it into the candidate set $C$ together with the candidates read from other servers. After the client receives $S-t$ candidates from different servers, \textsc{collect} returns.

In \textsc{filter}, the client submits $C$ to each server. Upon reception of $C$, the server chooses a candidate $c_{wb}$ to write-back, as the candidate with the highest timestamp in $C$ such that \textsf{valid}($c_{wb}$) holds, and sets $lc$ to $c_{wb}$ if $c_{wb}.ts > lc.ts$. Roughly speaking, the predicate \textsf{valid}($c$) holds if the server, verifies the integrity of the timestamp $c.ts$ and nonce $c.\nonce$ either by the MAC, or by the corresponding history entry. Besides that, the server chooses a candidate $c_{rt}$ to return, as the candidate with the highest timestamp in $C$ such that $\textsf{validByHist}(c_{rt})$ holds. The predicate \textsf{validByHist}($c$) holds, if the server keeps a matching history entry for $c$. The server then responds to the client with a message including the timestamp $c_{rt}.ts$, the fragment $Hist[c_{rt}.ts].fr$, the cross-checksum $Hist[c_{rt}.ts].cc$ and the vector of MACs $Hist[c_{rt}.ts].vec$.

The client waits for the responses from servers until there is a candidate $c$ with the highest timestamp in $C$ such that \textsf{safe}($c$) holds, or until $C$ is empty, after which \textsc{filter} returns. The predicate \textsf{safe}($c$) holds if at least $t+1$ different servers $s_i$ have responded with timestamp $c.ts$, fragment $fr_i$,  cross-checksum $cc$ such that $H(fr_i) = cc[i]$, and vector $vec$.
If $C$ is empty, the client sets $V$ to the initial value $\bot$. Otherwise, the client selects the highest candidate $c \in C$ and restores value $V$ by decoding $V$ from the $t+1$ correct fragments received for $c$.

In \textsc{repair}, the client verifies the integrity of $c.vec$ by matching it against the vector $vec$ received from $t+1$ different servers. If $c.vec$ and $vec$ match then \textsc{repair} returns. Otherwise, the client repairs $c$ by setting $c.vec$ to $vec$ and invokes a round of write-back by sending a \textsc{repair}$\langle tsr, c \rangle$ message to all servers. Upon reception of such a message, the server sets $lc$ to $c$ if $c.ts > lc.ts$ and \textsf{valid}$(c)$ holds and responds with an \textsc{repair\_ack} message to the client. Once the client receives acknowledgements from $n-t$ different servers, \textsc{repair} returns. After \textsc{repair} returns, the \textsc{READ} returns $V$.

\begin{algo}[t]
\small

\begin{distribalgo}[1] \setcounter{ALC@line}{\value{alg:client1:lines}}
\smallskip
\INDENT {\textbf{Definitions:}}
\STATE $tsr$: $num$, initially $0$
\STATE $Q,R$: set of $pid$, initially $\emptyset$
\STATE $C$: set of $(ts, \nonce, vec)$, initially $\emptyset$
\STATE $W[1 \dots S]$: vector of $(ts,fr,cc,vec)$, initially $[]$
\ENDINDENT
\setcounter{alg:client1:lines}{\value{ALC@line}}
\end{distribalgo}
\begin{tabular}{c}\hline\mbox{}\hspace{0.45\textwidth}\mbox{}\end{tabular}
\vspace{-2 em}
\begin{distribalgo}[1]   \setcounter{ALC@line}{\value{alg:client1:lines}}
\INDENT {\textbf{operation} \textsc{READ}$()$}
\STATE $C,Q,R \leftarrow \emptyset$
\STATE $tsr \leftarrow tsr + 1$
\STATE $C \leftarrow \textsc{collect}(tsr)$
\STATE $C \leftarrow \textsc{filter}(tsr, C)$
\IF {$C \neq \emptyset$}
\STATE $c \leftarrow c' \in C: \textsf{highcand}(c') \wedge \textsf{safe}(c')$~\label{alg2:reader:select}
\STATE $V \leftarrow \textsc{restore}(c.ts)$ ~\label{alg2:reader:restore}
\STATE $\textsc{repair}(c)$
\ENDIF
\STATE \textbf{else} $V \leftarrow \bot$
\STATE return $V$
\ENDINDENT

\smallskip
\INDENT {\textbf{procedure} \textsc{collect}($tsr$)}
\STATE send \textsc{collect}$\langle tsr \rangle$ to all servers
\STATE \textbf{wait until} $|Q| \geq S-t$
\STATE return $C$
\ENDINDENT

\smallskip
\INDENT {\textbf{upon} receiving \textsc{collect\_ack}$\langle tsr, c_i\rangle$ from server $s_i$}
\STATE $Q \leftarrow Q \cup \{ i \}$
\STATE \textbf{if} $c_i.ts > ts_0$ \textbf{then} $C \leftarrow C \cup \{c_i\}$
\ENDINDENT

\smallskip
\INDENT {\textbf{procedure} \textsc{filter}($tsr, C$)}
\STATE send \textsc{filter}$\langle tsr, C \rangle$ to all servers
\STATE \textbf{wait until} $|R| \geq S-t\ \wedge$\\ $((\exists c \in C: \textsf{highcand}(c) \wedge \textsf{safe}(c)) \vee C = \emptyset)$
\STATE return $C$
\ENDINDENT

\smallskip
\INDENT {\textbf{upon} receiving \textsc{filter\_ack}$\langle$$ $$tsr$$,$$ $$ts$$, $$fr$$, $$cc$$, $$vec$$\rangle$ from server $s_i$}
\STATE $R \leftarrow R \cup \{ i \}$; $W[i] \leftarrow (ts,fr,cc,vec)$
\STATE $C \leftarrow C \setminus \{c \in C: \textsf{invalid}(c) \}$
\ENDINDENT

\smallskip
\INDENT {\textbf{procedure} \textsc{restore}($ts$)}
\STATE $cc \leftarrow cc'$ s.t. $\exists R' \subseteq R : |R'| \geq t+1\ \bigwedge$\\ $(\forall i \in R' : W[i].ts = ts \wedge W[i].cc = cc')$ \label{alg2:reader:cc}
\STATE $F \leftarrow \{W[i].fr : i $$\in$$ R \wedge W[i].ts $$=$$ ts \wedge H(W[i].fr) $$=$$ cc[i]\}$ \label{alg2:reader:fr}
\STATE $V \leftarrow \mathrm{decode}(F, t+1, S)$ \label{alg2:reader:decode}
\STATE return $V$
\ENDINDENT

\smallskip
\INDENT {\textbf{procedure} \textsc{repair}($c$)}
\STATE $vec \leftarrow vec'$ s.t. $\exists R' \subseteq R : |R'| \geq t+1\ \bigwedge$\\ $(\forall i \in R' : W[i].ts = c.ts \wedge W[i].vec = vec')$
\IF {$c.vec \neq vec$} \label{alg2:reader:vec-integrity}
\STATE $c.vec \leftarrow vec$ \label{alg2:reader:repair}\hfill //repair
\STATE send \textsc{repair}$\langle tsr, c \rangle$ to all servers
\STATE \textbf{wait for} \textsc{repair\_ack}$\langle tsr \rangle$ from $S-t$ servers
\ENDIF
\ENDINDENT

\smallskip
\INDENT {\textbf{Predicates:}}
\STATE \textsf{highcand}($c$) $\triangleq$ ($c.ts = \max(\{c'.ts : c' \in C\}))$
\smallskip
\STATE \textsf{safe}($c$) $\triangleq$ $\exists R' \subseteq R: |R'| \geq t+1 \ \bigwedge$\\ \label{alg2:reader:safe}
$(\forall i \in R': W[i].ts = c.ts)\ \bigwedge$\\
$(\forall i,j \in R'$$:$$W[i].cc $$=$$ W[j].cc \wedge H(W[i].fr) $$=$$ W[j].cc[i])\bigwedge$\\
$(\forall i,j \in R': W[i].vec = W[j].vec$)
\smallskip
\STATE \textsf{invalid}($c$) $\triangleq$ $|\{i \in R: W[i].ts < c.ts \}| \geq S-t$
\ENDINDENT
\setcounter{alg:client1:lines}{\value{ALC@line}}
\end{distribalgo}
\begin{tabular}{c}\hline\mbox{}\hspace{0.45\textwidth}\mbox{}\end{tabular}
\caption{{Algorithm of client $r$ in \mprotocol.}}\label{alg2:reader}
\vspace{-1 em}
\end{algo}

\subsection{Correctness Sketch}
We show that \mprotocol\ satisfies \textsc{READ} linearizability as follows. We show that if a completed \textsc{READ} $rd$ returns $V$ then a subsequent \textsc{READ} $rd'$ does not return a value older than $V$. Note that the arguments for \textsc{READ}/\textsc{WRITE} linearizability and wait-freedom are very similar to those of \protocol{} (Section~\ref{sec:pow-sketch}), and therefore omitted.

Suppose a \textsc{READ} $rd'$ by a correct client follows after $rd$ that returns $V$. If $c$ is the candidate selected in $rd$, we argue that $rd'$ does not select a candidate with a timestamp lower than $c.ts$. Note that besides $c$ being a valid candidate, in \textsc{repair}, the client also checks the integrity of $c.vec$. If $c.vec$ passes the integrity check in $rd$ (line~\ref{alg2:reader:vec-integrity} of Algorithm~\ref{alg2:reader}), then the integrity of $c$ has been fully established. Otherwise, $c.vec$ fails the integrity check. In that case the client repairs $c$ (line~\ref{alg2:reader:repair} of Algorithm~\ref{alg2:reader}) and subsequently writes-back $c$ to $t+1$ correct servers. In both cases, $t+1$ correct servers have set $lc$ to $c$ or to a valid candidate with a higher timestamp. As such, during \textsc{collect} in $rd'$, a valid candidate $c'$ such that $c'.ts \geq c.ts$ is included in $C$. Since $c'$ is valid, $t+1$ correct servers hold history entries matching $c'$ and none of them responds with a timestamp lower than $c'.ts$. Consequently, at most $2t$ timestamps received by the client in \textsc{filter} are lower than $c'.ts$ and thus $c'$ is never excluded from $C$. By Algorithm~\ref{alg2:reader}, $rd'$ does not select a candidate with timestamp lower than $c'.ts \geq c.ts$.

\section{Implementation \& Evaluation}\label{sec:implementation}

In this section, we describe an implementation modeling a Key-Value Store (KVS) based on \mprotocol. More specifically, to model a KVS, we use multiple instances of \mprotocol\ referenced by keys. We then evaluate the performance of our implementation  and we compare it both: \emph{(i)} M-ABD, the multi-writer variant of the crash-only atomic storage protocol of~\cite{ABND95}, and \emph{(ii)} Phalanx, the multi-writer robust atomic protocol of~\cite{Phalanx}.

\subsection{Implementation Setup}\label{subsec:setup}

Our KVS implementation is based on the JAVA-based framework Neko~\cite{Neko09} that provides support for inter-process communication, and on the Jerasure~\cite{Jerasure08} library for constructing the dispersal codes.
To evaluate the performance of our \mprotocol\, we additionally implemented two KVSs  based on M-ABD and Phalanx.

In our implementation, we relied on 160-bit SHA1 for hashing purposes, 160-bit keyed HMACs to implement MACs, and 1024-bit DSA to generate signatures. For simplicity, we abstract away the effect of message authentication in our implementations; we argue that this does not affect our performance evaluation since data origin authentication is typically handled as part of the access control layer in all three implementations, when deployed in realistic settings (e.g., in Wide Area Networks).

We deployed our implementations on a private network consisting of a 4-core Intel Xeon E5443 with 4GB of RAM, a 4 core Intel i5-3470 with 8 GB of RAM,
an 8 Intel-Core i7-37708 with 8 GB of RAM, and a 64-core Intel Xeon E5606 equipped with 32GB of RAM. In our network, the communication between various machines was bridged
using a 1 Gbps switch. All the servers were running in separate processes on the Xeon E5606 machine, whereas the clients were distributed among the Xeon E5443, i7, and the i5 machines. Each client invokes operation in a closed loop, i.e., a client may have at most one pending operation. In all KVS implementations,  all \textsc{WRITE} and \textsc{READ} operations are served by a local database stored on disk.
%

We evaluate the peak throughput incurred in \mprotocol\ in \textsc{WRITE} and \textsc{READ} operations, when compared to M-ABD and Phalanx with respect to: \emph{(i)} the file size (64 KB, 128 KB, 256 KB, 512 KB, and 1024 KB), and \emph{(ii)}
to the server failure threshold $t$ (1, 2, and 3, respectively). For better evaluation purposes, we vary each variable separately and we fix
the remaining parameters to a default configuration (Table~\ref{tab:nominal}).
We also evaluate the latency incurred in \mprotocol\ with respect to the attained throughput.



\begin{table}
\centering
\scalebox{1.1}{\begin{tabular}{|c|c|}
  \hline
  \textbf{Parameter} & \textbf{Default Value}\\
  \hline
  \hline
Failure threshold $t$                          & 1 \\
File size                                & 256 KB\\
Probability of Concurrency & 1\%\\
Workload Distribution & 100\% \textsc{READ} /100\% \textsc{WRITE}\\
  \hline
\end{tabular}}
\caption{Default parameters used in evaluation.}
\vspace{-1 em}
\label{tab:nominal}
\end{table}

We measure peak throughput as follows. We require that each writer performs back to back \textsc{WRITE} operations; we then increase the number of writers in the system until
the aggregated throughput attained by all writers is saturated. The peak throughput is then computed as the maximum aggregated amount of data (in bytes) that can be written/read to the servers per second.

Each data point in our plots is averaged over 5 independent measurements; where appropriate, we include the corresponding 95\% confidence intervals. 
as data objects. On the other hand, \textsc{READ} operations request the data pertaining to randomly-chosen keys.
For completeness, we performed our evaluation \emph{(i)} in the Local Area Network (LAN) setting comprising our aforementioned network (Section~\ref{subsec:lan}) and \emph{(ii)} in a simulated Wide Area Network (WAN) setting (Section~\ref{subsec:wan}). Our evaluation results are presented in Figure~\ref{fig:results}.

\subsection{Evaluation Results within a LAN Setting}\label{subsec:lan}

\begin{figure*}[tb]
  \centering
   \subfigure[Throughput vs. latency in a LAN setting.
     ]{
          \label{fig:latency}
          \includegraphics*[width=0.47\linewidth]{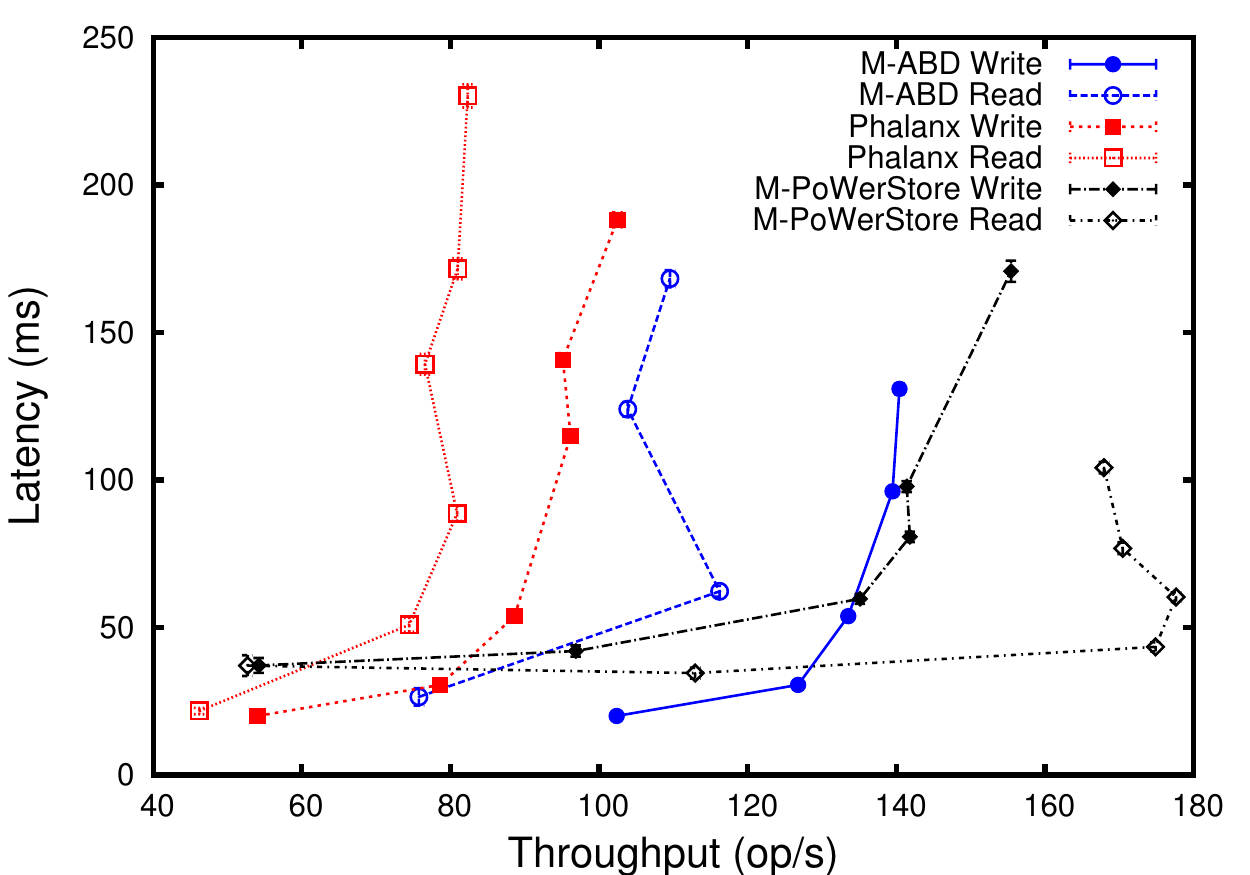}}
           \hspace{.01in}
     \subfigure[Peak throughput with respect to the failure threshold in a LAN setting.
     ]{
          \label{fig:newfaulty}
          \includegraphics*[width=0.47\linewidth]{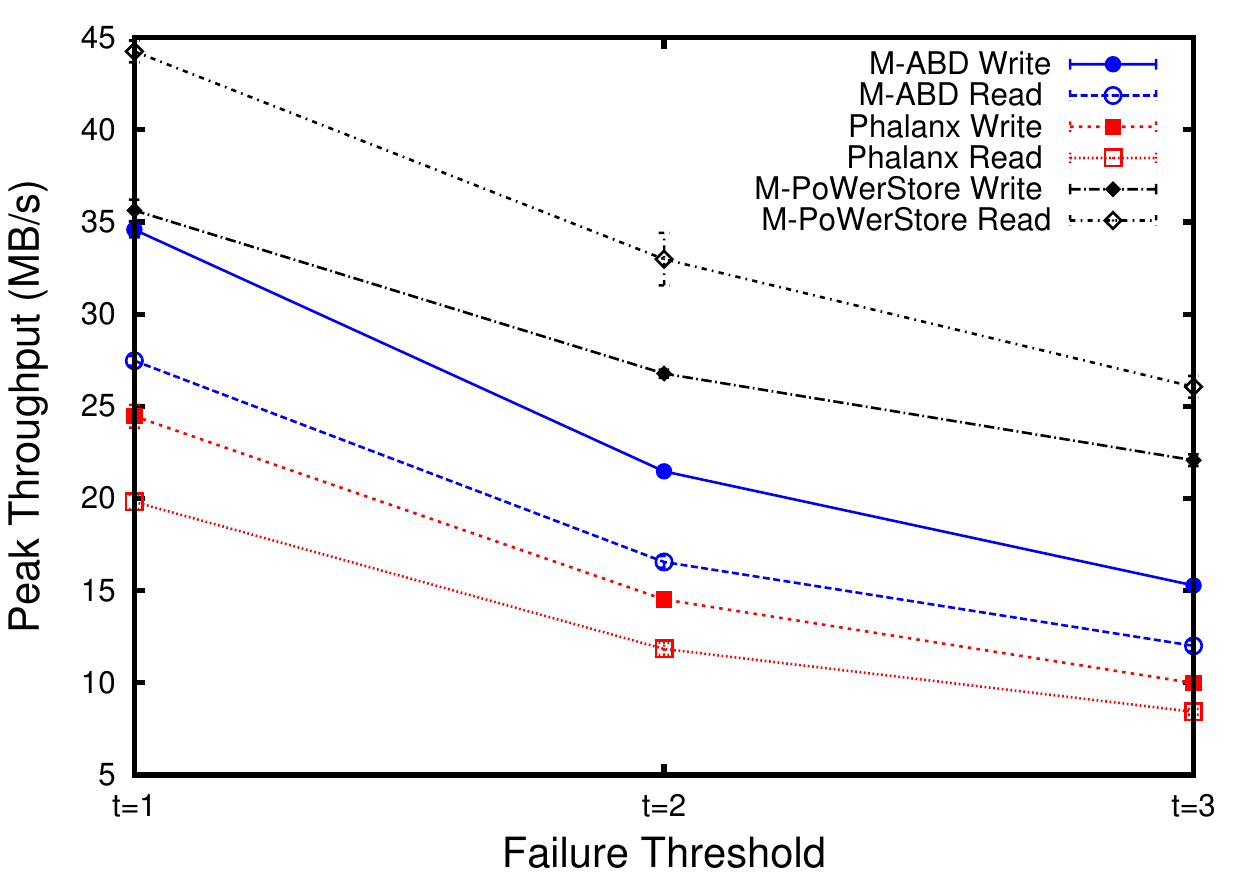}}\\
     \hspace{.01in}
     \subfigure[Peak throughput with respect to the file size in a LAN setting.
     ]{
         \label{fig:newfile}
         \includegraphics*[width=0.47\linewidth]{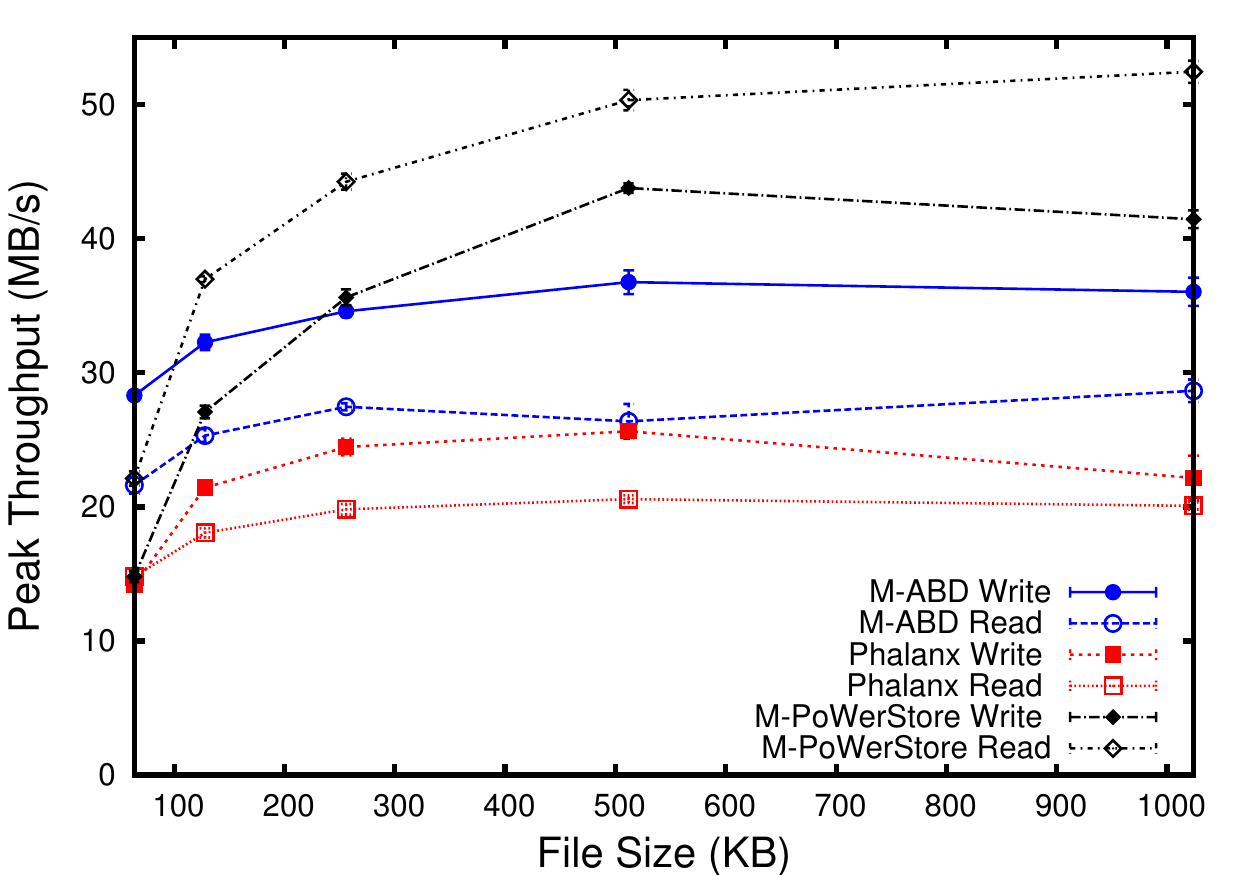}}
     \subfigure[Latency vs the failure threshold in a simulated WAN setting.
     ]{
          \label{fig:latencypareto}
          \includegraphics*[width=0.47\linewidth]{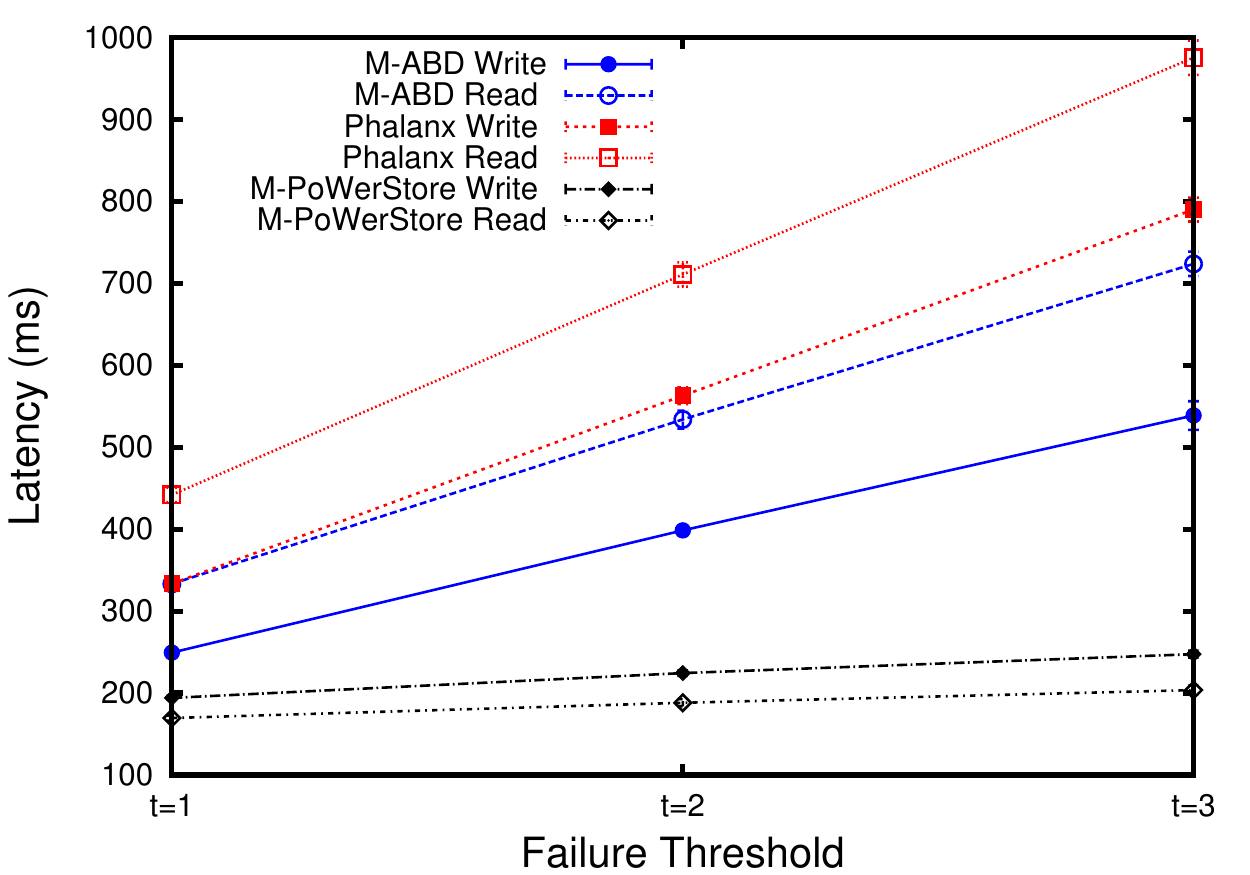}}
     \caption{Evaluation Results. Data points are averaged over 5 independent runs; where appropriate, we include the corresponding 95\% confidence intervals.}
    \label{fig:results}
\end{figure*}

Figure~\ref{fig:latency} depicts the latency incurred in \mprotocol{} when compared to M-ABD and Phalanx, with respect to the achieved throughput (measured in the number of operations per second).
Our results show that, by combining metadata write-backs with erasure coding, \mprotocol{} achieves lower latencies than M-ABD and Phalanx for both \textsc{READ} and \textsc{WRITE} operations.
As expected, \textsc{READ} latencies incurred in \protocol{} are lower than those of \textsc{WRITE} operations since a \textsc{WRITE} requires an extra communication round corresponding to the \textsc{clock} round. Furthermore, due to PoW and the use of lightweight cryptographic primitives, the \textsc{READ} performance of \protocol{} considerably outperforms M-ABD and Phalanx. On the other hand, writing in \mprotocol{} compares similarly to the writing in M-ABD.

Figure~\ref{fig:newfaulty} depicts the peak throughput achieved in \mprotocol{} with respect to the number of Byzantine servers $t$.
As $t$ increases, the gain in peak throughout achieved in \mprotocol's \textsc{READ} and \textsc{WRITE} increases compared to M-ABD and Phalanx. This mainly stems from the reliance on erasure coding, which ensures that
the overhead of file replication among the servers is minimized when compared to M-ABD and Phalanx. In typical settings, featuring $t=1$ and the default parameters of Table~\ref{tab:nominal}, the peak throughput achieved in \mprotocol's \textsc{READ} operation is almost twice as large as that in M-ABD and Phalanx.

In Figure~\ref{fig:newfile}, we measure the peak throughout achieved in \mprotocol{} with respect to the file size. Our findings clearly show that as the file size increases, the performance gain of \mprotocol{} compared to M-ABD and Phalanx becomes considerable. For example, when the file size is 1 MB, the peak throughput of \textsc{READ} and \textsc{WRITE} operations in \mprotocol{} approaches the (network-limited) bounds of 50 MB/s\footnote{Note that an effective peak throughout of 50MB/s in \mprotocol{} reflects an actual throughput of almost 820 Mbps when $t=1$.} and 45 MB/s, respectively.

\subsection{Evaluation Results within a Simulated WAN Setting}\label{subsec:wan}

We now proceed to evaluate the  performance of \mprotocol{} when deployed in WAN settings. For that purpose, we rely on a 100 Mbps switch to bridge the network outlined in Section~\ref{subsec:setup} and
we rely on NetEm~\cite{netem} to emulate the packet delay variance specific to WANs. More specifically, we add a Pareto distribution to our links, with a mean of 20 ms and a variance of 4 ms.

We then measure the latency incurred in \mprotocol{} in the emulated WAN setting. Our measurement results (Figure~\ref{fig:latencypareto}) confirm our previous analysis conducted in the LAN scenario and demonstrate the superior performance of \mprotocol{} compared to M-ABD and Phalanx in realistic settings. Here, we point out that the performance of \mprotocol{} incurred in both \textsc{READ} and \textsc{WRITE} operations does not deteriorate as the number of Byzantine servers in the system increases. This is mainly due to the reliance on erasure coding. In fact, the overhead of transmitting an erasure-coded file $F$ to the $3t+1$ servers, with a reconstruction threshold
of $t+1$ is given by $\frac{3t+1}{t+1}|F|$. It is easy to see that, as $t$ increases, this overhead is asymptotically increases towards $3|F|$.

\section{Related Work}\label{sec:relwork}

A seminal crash-tolerant \emph{robust} \emph{linearizable} read/write storage implementation assuming a majority of correct processes was presented in \cite{ABND95}. In the original single-writer variant of \cite{ABND95},  read operations always take 2 rounds between a client and servers with \emph{data} write-backs in the second round. On the other hand all write operations complete in a single round; in the multi-writer variant \cite{LS02}, the second write round is necessary. Server state modifications by readers introduced by \cite{ABND95} are unavoidable; namely, \cite{FL03} showed a $t+1$ lower bound on the number of servers that a reader has to modify in any wait-free linearizable storage. However, robust storage implementations differ in the strategy employed by readers: in some protocols readers write-back data (e.g., \cite{ABND95,Phalanx,AAB07,GNS09,GV10,DGM11}) whereas in others readers only write metadata to servers (e.g., \cite{FL03,DGLV10,CA12}).

The only two asynchronous storage protocols that feature \emph{metadata write-backs} are multi-writer crash-tolerant protocols of \cite{FL03} and \cite{CA12} that are both also linearizable, wait-free and optimally resilient. The read/write protocol of \cite{FL03} fails to achieve optimal latency featuring 3-round writes and reads. Vivace \cite{CA12} is a key-value storage system tailored for WANs (geo-replication); it features 3-round reads and 2-round writes, saving on a communication round by relying on synchronized clocks (NTP, GPS), which are used as counters. In comparison, \protocol\ features optimal latency without synchronized clocks and is the first protocol to implement metadata write-backs while tolerating Byzantine failures.

Data write-backs are also not needed in case of \emph{fast} robust storage implementations that feature single round reads and writes \cite{DGLV10}. Namely, \cite{DGLV10} presents fast single-writer crash-tolerant and BFT storage implementations in which readers only write metadata \emph{while} reading data in the single round of read and hence, without any write-back. However, fast implementations are fundamentally limited and cannot be optimally resilient, since the number of required servers is inherently linear in number of readers \cite{DGLV10}. The limitation on the number of readers of \cite{DGLV10} was relaxed in~\cite{GNS09}, where a single-writer crash-tolerant robust linearizable storage implementation was presented, in which most of the reads complete in a single round, yet a fraction of reads is permitted to be ``slow'' and complete in 2 rounds.

Clearly, most BFT storage implementations have been focusing on using as few servers as possible, ideally $3t+1$, which defines optimal resilience in the Byzantine model \cite{MAD02}. This is achieved by Phalanx \cite{Phalanx}, a BFT variant of \cite{ABND95}. Phalanx uses digital signatures, i.e., \emph{self-verifying data}, to port \cite{ABND95} to the Byzantine model, maintaining the latency of~\cite{ABND95}, as well as its data write-backs. However, since digital signatures introduce considerable overhead~\cite{Rei94,MR97}, research attention has shifted from protocols that employ self-verifying data \cite{Phalanx, MR98, CT06, LR06} to protocols that feature lightweight authentication, or no data authentication at all (unauthenticated model).

In the unauthenticated model,~\cite{ACKM06} ruled out the existence of optimally resilient robust Byzantine fault-tolerant storage implementation where all write operations finish after a single communication round. This explained the previous difficulties in reaching optimal resilience in unauthenticated BFT storage implementations where several protocols have used $4t+1$ servers \cite{GWGR04,BD04}. Furthermore,~\cite{DGM11} showed the impossibility of reading from a robust optimally resilient linearizable storage in two communication rounds; in addition, if \textsc{WRITE} operations perform a constant number of rounds, even reading in three rounds is impossible~\cite{DGM11}. These results imply that the optimal latency of a robust optimally resilient and linearizable BFT storage in the unauthenticated model is 2 rounds for writes and 4 rounds for reads, even in the single writer case. This can be achieved by the regular-to-linearizable transformation of the regular \cite{Lam86} storage protocol of \cite{GV06}. Hence, it is not surprising that other robust BFT storage protocols in the unauthenticated model focused on optimizing common-case latency with either an unbounded number of read rounds in the worst case \cite{GLV06,GV10} or a number of read rounds dependent on the number of faulty processes $t$ \cite{MAD02,AAB07}.

Clearly, there is a big gap between storage protocols that use self-verifying data and those that assume no authentication. Loft \cite{HendricksGR07} aims at bridging this gap and implements erasure-coded optimally resilient linearizable storage  while optimizing the failure-free case. Loft uses homomorphic fingerprints and MACs; it features 3-round wait-free writes, but reads are based on data write-backs and are only obstruction-free \cite{HLM03}, i.e., the number of read rounds is unbounded in case of read/write concurrency. Similarly, our \emph{Proofs of Writing} (PoW) incorporate lightweight authentication that is, however, sufficient to achieve optimal latency and to facilitate metadata write-backs. We find PoW to be a fundamental improvement in the light of BFT storage implementations that explicitly renounce linearizability in favor of weaker regularity due to the high cost of data write-backs \cite{BessaniCQAS11}.

\section{Concluding Remarks}
\label{sec:conclusion}

In this paper, we presented \protocol, the first efficient
robust storage protocol that achieves optimal latency, measured
in maximum (worst-case) number of communication rounds between a client and storage servers, without resorting to digital signatures.
We also separately presented a multi-writer variant of our protocol called \mprotocol.  The \emph{efficiency} of our proposals stems from combining \emph{lightweight cryptography}, \emph{erasure coding} and  \emph{metadata writebacks}, where readers write-back only metadata to achieve linearizability. While robust BFTs have been often criticized for being prohibitively inefficient, our findings suggest that efficient and robust BFTs can be realized
in practice by relying on lightweight cryptographic primitives without compromising \emph{worst-case} performance.

At the heart of both \protocol\ and \mprotocol{} protocols are \emph{Proofs of Writing (PoW)}: a novel \scheme\ inspired by commitment schemes in the flavor of~\cite{Halevi1996}, that enables \protocol\ to write and read in 2 rounds in case of the single-writer \protocol\ which we show optimal.
Similarly, by relying on PoW, \mprotocol{} features 3-round writes/reads where the third read round is only invoked under active attacks.
Finally, we  demonstrated \mprotocol's superior performance compared to existing crash and Byzantine-tolerant atomic storage implementations.

We point out that our protocols assume unbounded storage capacity to maintain a version history of
the various updates performed by the writers in the system. We argue, however, that this limitation is not particular to our proposals and is inherent to all
solutions that rely on erasure-coded data~\cite{Soules2003}. Note that prior studies have demonstrated that it takes several weeks to exhaust the capacity of versioning systems~\cite{Strunk2000}; in case the storage capacity is bounded, the servers can rely on efficient garbage collection mechanisms~\cite{GWGR04} to avoid possible exhaustion of the storage system.

\newpage

\appendix

\section{Correctness Proofs}
\subsection{Correctness of \protocol}

\begin{defn}[Valid candidate] \label{def:validcand}
A candidate $c$ is \emph{valid} if \emph{\textsf{valid}}$(c)$ holds at some correct server.
\end{defn}

\begin{defn}[Timestamps of operations]
A \textsc{read} operation $rd$ by a correct reader has timestamp $ts$ iff the reader in $rd$ selected $c$
in line~\ref{alg:reader:select} such that $c.ts = ts$. A \textsc{write} operation $wr$ has timestamp $ts$ iff
the writer increments its timestamp to $ts$ in line~\ref{alg:writer:timestamp}.
\end{defn}

\begin{la}[Validity]\label{la:validity}
Let $rd$ be a completed \textsc{read} by a correct reader. If $rd$ returns value $V \neq \bot$ then $V$ was written.
\end{la}
\begin{prooff} We show that if $V$ is the value decoded in line~\ref{alg:reader:decode}, then $V$ was indeed written. To show this, we argue that the fragments used to decode $V$ were written. Note that prior to decoding $V$ from a set of fragments, the reader establishes the correctness of each fragment as follows. First, in line~\ref{alg:reader:cc}, the reader chooses a cross-checksum that was received from $t+1$ servers. Since one of these servers is correct, the chosen cross-checksum was indeed written. Secondly, the reader checks in line~\ref{alg:reader:fr} that each of the $t+1$ fragments used to decode $V$ hashes to the corresponding entry in the cross-checksum.  By the collision-resistance of $H$, all fragments that pass this check were indeed written. Therefore, if $V$ is the value decoded from these fragments, we conclude that $V$ was written.
\end{prooff}

\begin{la}[Proofs of Writing] \label{la:pow}
If $c$ is a valid candidate, then there exists a set $Q$ of $t+1$ correct servers such that each server $s_i \in Q$ changed $Hist[c.ts]$ to $(fr_i,cc,H(c.N))$.
\end{la}
\begin{prooff}
If $c$ is valid, then by Definition~\ref{def:validcand}, \textsf{valid}($c$) is true at some correct server $s_j$. Hence, $H(c.N) = Hist[c.ts].N_H$ holds at $s_j$. By the pre-image resistance of $H$, no computationally bounded adversary can acquire $c.N$ from the sole knowledge of $H(c.N)$. Hence, $c.N$ stems from the writer in a \textsc{write} operation $wr$ with timestamp $c.ts$. By Algorithm~\ref{alg:writer}, line~\ref{alg:writer:complete}, the value of $c.N$ is revealed after the \textsc{store} phase in $wr$ completed. Hence, there exists a set $Q$ of $t+1$ correct servers such that each server $s_i \in Q$ changed $Hist[c.ts]$ to $(fr_i,cc,H(c.N))$.
\end{prooff}

\begin{la}[No exclusion]\label{la:noexclusion}
Let $c$ be a valid candidate and let $rd$ be a \textsc{read} by a correct reader that includes $c$ in $C$ during \textsc{collect}. Then $c$ is never excluded from $C$.
\end{la}
\begin{prooff}
As $c$ is valid, by Lemma~\ref{la:pow} a there exists a set $Q$ of $t+1$ correct servers such that each server $s_i \in Q$ changed $Hist[c.ts]$ to $(*,*,H(c.N))$. Hence, \textsf{valid}($c$) is true at every server in $Q$. Thus, no server in $Q$ replies with a timestamp $ts < c.ts$ in line~\ref{alg:server:ret}.
Therefore, at most $S-t-1= 2t$ timestamps received by the reader in the \textsc{filter} phase are lower than $c.ts$, and so $c$ is never excluded from $C$.
\end{prooff}

\begin{la}[\textsc{read}/\textsc{write} Atomicity]~\label{la:rwatomic}
Let $rd$ be a completed \textsc{read} by a correct reader. If $rd$ follows some complete \textsc{write}$(V)$, then $rd$ does not return a value older than $V$.
\end{la}
\begin{prooff}
If $ts$ is the timestamp of \textsc{write}$(V)$, it is sufficient to show that the timestamp of $rd$ is not lower than $ts$. To prove this, we show that $\exists c' \in C$ such that \emph{(i)} $c'.ts \geq ts$ and \emph{(ii)} $c'$ is never excluded from $C$.

By the time $\textsc{write}(V)$ completes, $t+1$ correct servers hold in $lc$ a candidate whose timestamp is $ts$ or greater. According to lines~\ref{alg:server:update-complete},~\ref{alg:server:update-gc} of Algorithm~\ref{alg:server}, a correct server never changes $lc$ to a candidate with a lower timestamp. Hence, when $rd$ is invoked, $t+1$ correct servers hold candidates with timestamp $ts$ or greater in $lc$. Hence, during the \textsc{collect} phase in $rd$, some candidate received from a correct server with timestamp $ts$ or greater is inserted in $C$. Such a candidate is necessarily valid because either the server received it directly from the writer, or the server checked its integrity in line~\ref{alg:server:valid-check}. Let $c'$ be the valid candidate with the highest timestamp in $C$. Then by Lemma~\ref{la:noexclusion}, $c'$ is never excluded from $C$. By line~\ref{alg:reader:select}, no candidate $c$ such that $c.ts < c'.ts$ is selected. Since $c'.ts \geq ts$, no candidate with a timestamp lower than $ts$ is selected in $rd$.
\end{prooff}

\begin{la}[\textsc{read} atomicity]\label{la:ratomic}
Let $rd$ and $rd'$ be two completed \text{read} operations by correct readers. If $rd'$ follows $rd$ that returns $V$, then $rd'$ does not return a value older than $V$.
\end{la}
\begin{prooff} If $c$ is the candidate selected in $rd$, it is sufficient to show that the timestamp of $rd'$ is not lower than $c.ts$. We argue that
$C$ contains a candidate $c'$ such that \textit{(i)} $c'.ts \geq c.ts$ and \textit{(ii)} $c'$ is never excluded from $C$.

By the time $rd$ completes, $t+1$ correct servers hold $c$ in $LC$. As $c$ was selected in $rd$ in line~\ref{alg:reader:select}, some correct server asserted that $c$ is valid in line~\ref{alg:server:valid-check}. According to Algorithm~\ref{alg:server}, if a correct server excludes $c$ from $LC$ in line~\ref{alg:server:gc}, then the server changed $lc$ to a valid candidate with timestamp $c.ts$ or greater in line~\ref{alg:server:update-gc}. Consequently, $t+1$ correct servers hold in $LC \cup \{ lc \}$ a valid candidate with timestamp $c.ts$ or greater. As such, during \textsc{collect} in $rd'$, a valid candidate $c'$ such that $c'.ts \geq c.ts$ is included in $C$, and by Lemma~\ref{la:noexclusion}, $c'$ is never excluded from $C$. By line~\ref{alg:reader:select}, no candidate with a timestamp lower than $c'.ts$ is selected. Since $c'.ts \geq c.ts$, no candidate with a timestamp lower than $c.ts$ is selected in $rd'$.
\end{prooff}

\begin{theo}[Atomicity]
Algorithms~\ref{alg:writer},~\ref{alg:server} and~\ref{alg:reader} are atomic.
\end{theo}
\begin{prooff} This proof follows directly from Lemmas~\ref{la:validity}~\ref{la:rwatomic}~\ref{la:ratomic}.
\end{prooff}

We now proceed to proving wait-freedom.
\begin{theo}[Wait-freedom] Algorithms~\ref{alg:writer},~\ref{alg:server} and~\ref{alg:reader} are wait-free.
\end{theo}
\begin{prooff} We show that no operation invoked by a correct client ever blocks. The wait-freedom argument of the \textsc{write} is straightforward; in every phase, the writer awaits acks from the least number $S-t$ of correct servers. The same argument holds for the \textsc{collect} phase of the \textsc{read}. Hence, in the remainder of the proof, we show that no \textsc{read} blocks in the \textsc{filter} phase. By contradiction, consider a \textsc{read} $rd$ by reader $r$ that blocks during the \textsc{filter} phase after receiving \textsc{filter\_ack} messages from all correct servers. We distinguish two cases: (Case 1) $C$ includes a valid candidate and (Case 2) $C$ includes no valid candidate.

\begin{itemize}
\item Case 1: Let $c$ be the highest valid candidate included in $C$. We show that \textsf{highcand}($c$) $\wedge$ \textsf{safe}($c$) holds. Since $c$ is valid, by Lemma~\ref{la:pow}, there exists a set $Q$ of $t+1$ correct servers such that each server $s_i \in Q$ changed $Hist[c.ts]$ to $(fr_i,cc,H(c.N))$. Thus, during the \textsc{filter} phase, \textsf{valid}($c$) holds at every server in $Q$. As no valid candidate in $C$ has a higher timestamp than $c$, \textit{(i)} all servers $s_i \in Q$ (at least $t+1$) responded with timestamp $c.ts$, corresponding erasure coded fragment $fr_i$, cross-checksum $cc$ in line~\ref{alg:server:ret} and \textit{(ii)} all correct servers (at least $S-t$) responded with timestamps at most $c.ts$. By \emph{(i)}, $c$ is \textsf{safe}. By \emph{(ii)}, every $c' \in C$ such that $c'.ts > c.ts$ became \textsf{invalid} and was excluded from $C$, implying that $c$ is \textsf{highcand}.

\item Case 2: Here, we show that $C = \emptyset$. As none of the candidates in $C$ is valid, during the \textsc{filter} phase, the integrity check in line~\ref{alg:server:valid-check} failed for every candidate in $C$ at all correct servers. Hence, at least $S-t$ servers responded with timestamp $ts_0$. Since $ts_0$ is lower than any candidate timestamp, all candidates were classified as \textsf{invalid} and were excluded from $C$.
\end{itemize}
\end{prooff}

\begin{theo}[Latency]
Algorithms~\ref{alg:writer},~\ref{alg:server} and~\ref{alg:reader} feature a latency of \emph{two} communication rounds for the \textsc{write} and \emph{two} for the \textsc{read}.
\end{theo}
\begin{prooff} By Algorithm~\ref{alg:writer}, the \textsc{write} completes after two phases, \textsc{store} and \textsc{\complete}, each taking one communication round. By Algorithm~\ref{alg:reader}, the \textsc{read} completes after two phases, \textsc{collect} and \textsc{filter}, each incurring one communication round.
\end{prooff}

\subsection{Correctness of \mprotocol}

\begin{defn}[Valid candidate] \label{def2:validcand}
A candidate $c$ is \emph{valid} iff \emph{\textsf{valid}(c)} is true at some correct server.
\end{defn}

\begin{defn}[Timestamps of operations]
A \textsc{read} operation $rd$ by a non-malicious reader has timestamp $ts$ iff the reader in $rd$ selected $c$
in line~\ref{alg2:reader:select} such that $c.ts = ts$. A \textsc{write} operation $wr$ has timestamp $ts$ iff the \textsc{clock} procedure in $wr$ returned $ts$ in line~\ref{alg2:writer:clock}.
\end{defn}

\begin{la}[Validity]\label{la2:validity}
Let $rd$ be a completed \textsc{read} by a correct reader. If $rd$ returns value $V \neq \bot$ then $V$ was written.
\end{la}
\begin{prooff} We show that if $V$ is the value decoded in line~\ref{alg2:reader:decode}, then $V$ was indeed written. To show this, we argue that the fragments used to decode $V$ were written. Note that prior to decoding $V$ from a set of fragments, the reader establishes the correctness of each fragment as follows. First, in line~\ref{alg2:reader:cc}, the reader chooses a cross-checksum that was received from $t+1$ servers. Since one of these servers is correct, the chosen cross-checksum was indeed written. Secondly, the reader checks in line~\ref{alg2:reader:fr} that each of the $t+1$ fragments used to decode $V$ hashes to the corresponding entry in the cross-checksum.  By the collision-resistance of $H$, all fragments that pass this check were indeed written. Therefore, if $V$ is the value decoded from these fragments, we conclude that $V$ was written.
\end{prooff}

\begin{la}[\textsc{write} atomicity]\label{la2:powr}
Let $op$ be a completed operation by a correct client and let $wr$ be a completed $\textsc{write}$ such that $op$ precedes $wr$. If $ts_{op}$ and $ts_{wr}$ are the timestamps of $op$ and $wr$ respectively, then $ts_{wr}>ts_{op}$.
\end{la}
\begin{prooff} By the time $op$ completes, $t+1$ correct servers hold in $lc$ a candidate whose timestamp is $ts_{op}$ or greater. According to lines~\ref{alg2:server:update-complete},~\ref{alg2:server:update-filter},~\ref{alg2:server:update-repair} of Algorithm~\ref{alg2:server}, a correct server never updates $lc$ with a candidate that has a lower timestamp. Hence, the writer in $wr$ obtains from the \textsc{clock} procedure a timestamp that is greater or equal to $ts_{op}$ from some correct server $s_i$. Let $c$ be the candidate held in $lc$ by server $s_i$, and let $c.ts$ be the timestamp reported to the writer. We now argue that $c.ts$ is not fabricated. To see why, note that prior to overwriting $lc$ with $c$ in line~\ref{alg2:server:update-filter} (resp. ~\ref{alg2:server:update-repair}), server $s_i$ checks that $c$ is valid in line \ref{alg2:server:valid-filter} (resp. ~\ref{alg2:server:valid-repair}). The \textsf{valid} predicate as defined in line~\ref{alg2:server:valid-pred} subsumes an integrity check for $c.ts$. Hence, $c.ts$ passes the integrity check in line~\ref{alg2:writer:ts-integrity}; according to the \textsc{write} algorithm, $ts_{wr} \geq (c.ts.num+1, *, *) > c.ts \geq ts_{op}$.
\end{prooff}

\begin{la}[Proofs of Writing] \label{la2:pow}
If $c$ is a valid candidate, then there exists a set $Q$ of $t+1$ correct servers such that each server $s_i \in Q$ changed $Hist[c.ts]$ to $(fr_i,cc,H(c.N),vec)$.
\end{la}
\begin{prooff}
If $c$ is valid, then by Definition~\ref{def2:validcand}, \textsf{valid}($c$) is true at some correct server $s_j$. Hence, either $H(c.N) = Hist[c.ts].N_H$ or $\mathrm{verify}(c.vec[j], c.ts, H(c.N), k_j$) must hold at $s_j$. By the pre-image resistance of $H$, no computationally bounded adversary can acquire $c.N$ from the sole knowledge of $H(c.N)$. Hence, $c.N$ stems from some writer in a \textsc{write} operation $wr$ with timestamp $c.ts$. By Algorithm~\ref{alg2:writer}, line~\ref{alg2:writer:complete}, the value of $c.N$ is revealed after the \textsc{store} round in $wr$ completed. Hence, there exists a set $Q$ of $t+1$ correct servers such that each server $s_i \in Q$ changed $Hist[c.ts]$ to $(fr_i,cc,H(c.N),vec)$.
\end{prooff}

\begin{la}[No exclusion]\label{la2:noexclusion}
Let $c$ be a valid candidate and let $rd$ be a \textsc{read} by a correct reader that includes $c$ in $C$ during \textsc{collect}. Then $c$ is never excluded from $C$.
\end{la}
\begin{prooff}
As $c$ is valid, by Lemma~\ref{la2:pow} a there exists a set $Q$ of $t+1$ correct servers such that each server $s_i \in Q$ changed $Hist[c.ts]$ to $(*,*,H(c.N),vec)$. Hence, \textsf{validByHist}($c$) is true at every server in $Q$. Thus, no server in $Q$ replies with a timestamp $ts < c.ts$ in line~\ref{alg2:server:ret}.
Therefore, at most $S-t-1= 2t$ timestamps received by the reader in the \textsc{filter} round are lower than $c.ts$, and so $c$ is never excluded from $C$.
\end{prooff}

\begin{la}[\textsc{read}/\textsc{write} Atomicity]~\label{la2:rwatomic}
Let $rd$ be a completed \textsc{read} by a correct reader. If $rd$ follows some complete \textsc{write}$(V)$, then $rd$ does not return a value older than $V$.
\end{la}
\begin{prooff}
If $ts$ is the timestamp of \textsc{write}$(V)$, it is sufficient to show that the timestamp of $rd$ is not lower than $ts$. To prove this, we show that $\exists c' \in C$ such that \emph{(i)} $c'.ts \geq ts$ and \emph{(ii)} $c'$ is never excluded from $C$.

By the time $\textsc{write}(V)$ completes, $t+1$ correct servers hold in $lc$ a candidate whose timestamp is $ts$ or greater. According to lines~\ref{alg2:server:update-complete},~\ref{alg2:server:update-filter},~\ref{alg2:server:update-repair} of Algorithm~\ref{alg2:server}, a correct server never changes $lc$ to a candidate with a lower timestamp. Hence, when $rd$ is invoked, $t+1$ correct servers hold candidates with timestamp $ts$ or greater in $lc$. Hence, during \textsc{collect} in $rd$, some candidate received from a correct server with timestamp $ts$ or greater is inserted in $C$. Such a candidate is necessarily valid by the integrity checks in lines~\ref{alg2:server:valid-filter},~\ref{alg2:server:valid-repair}. Let $c'$ be the valid candidate with the highest timestamp in $C$. Then by Lemma~\ref{la2:noexclusion}, $c'$ is never excluded from $C$. By line~\ref{alg2:reader:select}, no candidate $c$ such that $c.ts < c'.ts$ is selected. Since $c'.ts \geq ts$, no candidate with a timestamp lower than $ts$ is selected in $rd$.
\end{prooff}

\begin{la}
\label{la2:rratomic}
\emph{(\textsc{read} atomicity).}
Let $rd$ and $rd'$ be two completed \text{read} operations by correct readers. If $rd'$ follows $rd$ that returns $V$, then $rd'$ does not return a value older than $V$.
\end{la}
\begin{prooff} If $c$ is the candidate selected in $rd$, it is sufficient to show that the timestamp of $rd'$ is not lower than $c.ts$. We argue that
$C$ contains a candidate $c'$ such that \textit{(i)} $c'.ts \geq c.ts$ and \textit{(ii)} $c'$ is never excluded from $C$.

As $c$ is selected in $rd$ in line~\ref{alg2:reader:select} only if \textsf{safe}($c$) holds, some correct server verified the integrity of $c.ts$ and $c.N$ in line~\ref{alg2:server:valid-nonce}. In addition, in \textsc{repair}, the reader in $rd$ checks the integrity of $c.vec$. We distinguish two cases:
\begin{itemize}
\item Case 1: If $c.vec$ passes the integrity check in line~\ref{alg2:reader:vec-integrity}, then the integrity of $c$ has been fully established. Hence, by the time $rd$ completes, $t+1$ correct servers validated $c$ in line~\ref{alg2:server:valid-filter} and changed $lc$ to $c$ or to a higher valid candidate.
\item Case 2: If vector $c.vec$ fails the integrity check in line~\ref{alg2:reader:vec-integrity}, then in \textsc{repair}, $c$ is repaired in line~\ref{alg2:reader:repair} and subsequently written back to $t+1$ correct servers. Hence, by the time $rd$ completes, $t+1$ correct servers validated $c$ in line~\ref{alg2:server:valid-repair} and changed $lc$ to $c$ or to a higher valid candidate.
\end{itemize}
Consequently, in the \textsc{collect} round in $rd'$ a valid candidate $c'$ such that $c'.ts \geq c$ is included in $C$, and by Lemma~\ref{la2:noexclusion}, $c'$ is never excluded from $C$. By line~\ref{alg2:reader:select}, no candidate with a timestamp lower than $c'$ is selected. Since $c'.ts \geq c.ts$, no candidate with a timestamp lower than $c.ts$ is selected in $rd'$.
\end{prooff}

\begin{theo}[Atomicity]
Algorithms~\ref{alg2:writer},~\ref{alg2:server} and~\ref{alg2:reader} are atomic.
\end{theo}
\begin{prooff} This proof follows directly from Lemmas~\ref{la2:validity}~\ref{la2:powr}~\ref{la2:rwatomic}~\ref{la2:rratomic}.
\end{prooff}

We now proceed to proving wait-freedom.
\begin{theo}[Wait-freedom] Algorithms~\ref{alg2:writer},~\ref{alg2:server},and~\ref{alg2:reader} are wait-free.
\end{theo}
\begin{prooff} We show that no operation invoked by a correct client ever blocks. The wait-freedom argument of the \textsc{write} is straightforward; in every round, the writer awaits acks from the least number $S-t$ of correct servers. The same argument holds for the \textsc{collect} and the \textsc{repair} rounds of the \textsc{read}. Hence, in the remainder of the proof, we show that no \textsc{read} blocks in the \textsc{filter} round. By contradiction, consider a \textsc{read} $rd$ by reader $r$ that blocks during the \textsc{filter} round after receiving \textsc{filter\_ack} messages from all correct servers. We distinguish two cases: (Case 1) $C$ includes a valid candidate and (Case 2) $C$ includes no valid candidate.

\begin{itemize}
\item Case 1: Let $c$ be the highest valid candidate included in $C$. We show that \textsf{highcand}($c$) $\wedge$ \textsf{safe}($c$) holds. Since $c$ is valid, by Lemma~\ref{la2:pow}, there exists a set $Q$ of $t+1$ correct servers such that each server $s_i \in Q$ changed $Hist[c.ts]$ to $(fr_i,cc,H(c.N),vec)$. Thus, during the \textsc{filter} round, \textsf{validByHist}($c$) holds at every server in $Q$. As no valid candidate in $C$ has a higher timestamp than $c$, \textit{(i)} all servers $s_i \in Q$ (at least $t+1$) responded with timestamp $c.ts$, corresponding erasure coded fragment $fr_i$, cross-checksum $cc$ and repair vector $vec$ in line~\ref{alg2:server:ret} and \textit{(ii)} all correct servers (at least $S-t$) responded with timestamps at most $c.ts$. By \emph{(i)}, $c$ is \textsf{safe}. By \emph{(ii)}, every $c' \in C$ such that $c'.ts > c.ts$ became \textsf{invalid} and was excluded from $C$, implying that $c$ is \textsf{highcand}.

\item Case 2: Here, we show that $C = \emptyset$. As none of the candidates in $C$ is valid, during \textsc{filter}, the integrity check in line~\ref{alg2:server:valid-nonce} failed for every candidate in $C$ at all correct servers. Hence, at least $S-t$ servers responded with timestamp $ts_0$. Since $ts_0$ is lower than any candidate timestamp, all candidates were classified as \textsf{invalid} and were excluded from $C$.
\end{itemize}
\end{prooff}

\begin{theo}[Non-skipping Timestamps]
Algorithms~\ref{alg2:writer},~\ref{alg2:server} and~\ref{alg2:reader} implement non-skipping timestamps.
\end{theo}
\begin{prooff} By construction, a fabricated timestamp would fail the check in line~\ref{alg2:writer:ts-integrity}. Hence, no fabricated timestamp is ever used in a \textsc{write}. The Lemma then directly follows from the algorithm of \textsc{write}.
\end{prooff}

\begin{theo}[Latency]
Algorithms~\ref{alg2:writer},~\ref{alg2:server} and~\ref{alg2:reader} feature a latency of \emph{three} communication rounds for the \textsc{write} and \emph{two} for the \textsc{read} in the absence of attacks. In the worst case, the \textsc{read} latency is \emph{three} communication rounds.
\end{theo}
\begin{prooff} By Algorithm~\ref{alg2:writer}, the \textsc{write} completes after three rounds, \textsc{clock}, \textsc{store} and \textsc{\complete}, each taking one communication round. In the absence of attacks, by Algorithm~\ref{alg2:reader}, the \textsc{read} completes after two rounds, \textsc{collect} and \textsc{filter}, each taking one communication round. Under BigMac~\cite{CWADM09} attacks the \textsc{read} may go to the \textsc{repair} round, incurring one additional communication round.
\end{prooff}

\end{document}